\documentclass[11pt,a4paper]{article}

\usepackage{jheppub}

\usepackage{amsmath}
\usepackage{amsthm}
\usepackage{amscd}
\usepackage{amsxtra}
\usepackage{upref}
\usepackage{amsfonts}
\usepackage{amssymb}
\usepackage{bbold}
\usepackage{graphicx}

\usepackage{pifont}

\DeclareMathOperator{\Tr}{Tr}

\numberwithin{equation}{section}

\title{U-duality in three and four dimensions}
\author{Emanuel Malek}
\affiliation{Department of Applied Mathematics and Theoretical Physics, \\Centre for Mathematical Sciences, University of Cambridge, \\ Wilberforce Road, Cambridge CB3 0WA, United Kingdom}
\emailAdd{E.Malek@damtp.cam.ac.uk}

\abstract{Using generalised geometry we study the action of U-duality acting in three and four dimensions on the bosonic fields of eleven dimensional supergravity. We compare the U-duality symmetry with the T-duality symmetry of double field theory and see how the $SL(2)\otimes SL(3)$ and $SL(5)$ U-duality groups reduce to the $SO(2,2)$ and $SO(3,3)$ T-duality symmetry groups of the type IIA theory. As examples we dualise M2-branes, both black and extreme. We find that uncharged black M2-branes become charged under U-duality, generalising the Harrison transformation, while extreme M2-branes will become new extreme M2-branes. The resulting tension and charges are quantised appropriately if we use the discrete U-duality group $E_d(Z)$.}


\begin{document}

\maketitle
\flushbottom

\newcommand{\lb}{\ensuremath{\mathopen{<}}}
\newcommand{\rb}{\ensuremath{\mathclose{>}}}
\newcommand{\ph}{\ensuremath{\phantom}}
\newcommand{\ov}{\ensuremath{\overline}}
\newcommand{\og}{\ensuremath{\bar{g}}}
\newcommand{\gm}{\ensuremath{\mathcal{H}}}

\newcommand{\be}{\begin{equation}}
\newcommand{\ee}{\end{equation}}

\newtheorem{defn}{Definition}[section]
\newtheorem{prop}[defn]{Proposition}
\newtheorem{thm}[defn]{Theorem}
\newtheorem{cor}[defn]{Corollary}
\newtheorem{claim}[defn]{Claim}
\newtheorem{rmk}[defn]{Remark}
\newtheorem{notation}[defn]{Notation}

\theoremstyle{definition}
\newtheorem{ex}[defn]{Example}

\section{Introduction} \label{SIntro}

By compactifying the 11- and 10-dimensional supergravities one obtains lower-dimensional theories exhibiting a group of global non-compact symmetries \cite{Cremmer:1977tt,Cremmer:1978ds,Cremmer:1979up}. These symmetries originate from the U- and T-dualities of the parent M- and string theory, respectively, and generate transformations linking different solutions of the supergravities through a web of dualities. The canonical example is a 10-dimensional solution of type IIA supergravity compactified on a circle of radius $R$ that is related by a Buscher T-duality transformation to a type IIB solution on a circle with inverse radius $1/R$ \cite{Buscher:1987sk,Buscher:1987qj}.

There is evidence that rather than being accidental symmetries arising only upon compactification, these dualities form the inherent symmetries of the 11-dimensional supergravity \cite{Nicolai:1986jk,Duff:1985bv,deWit:1986mz,deWit:2000wu,West:2010rv,West:2001as,West:2003fc,West:2004kb,Kleinschmidt:2003jf,West:2004iz,West:2011mm,Riccioni:2007ni,Nicolai:2010zz,Damour:2002cu,Hillmann:2009ci}. A natural formulation is generalised geometry \cite{Gualtieri:2003dx,Hitchin:2004ut,Hitchin:2005in,Hitchin:2005cv} which allows us to rewrite the bosonic part of the 11-dimensional supergravity action in a way which makes it manifestly invariant under the duality symmetries \cite{Hull:2007zu,Berman:2010is,Berman:2011pe,Berman:2011jh,Berman:2011cg}. This comes at the cost of introducing coordinates which are Fourier dual to winding modes of the M2- and M5-branes.\footnote{For example, the M2-brane has winding modes corresponding to an antisymmetric central charge in the duality algebra, $Z^{ij}$. This is interpreted as a momentum which is Fourier dual to a coordinate $y_{ij}$. See references \cite{Berman:2010is, Berman:2011jh} for M-theory and \cite{Hull:2009zb} for string theory.} A similar formulation has been developed for T-duality of 10-dimensional supergravities \cite{Hull:2009zb,Hull:2009mi,Hohm:2010pp,Hohm:2010jy}.\footnote{The reader is referred to references \cite{Hull:2006tp,Pacheco:2008ps,Aldazabal:2010ef,Coimbra:2011ky} for more about generalised geometry and M-theory and references \cite{Hull:2004in,Hull:2005hk,Dabholkar:2005ve,Hull:2006va,Grana:2008yw,Hull:2009sg,Aldazabal:2011nj,Grana:2012rr,Dibitetto:2012rk,Andriot:2012wx,Andriot:2012an,Jeon:2010rw,Jeon:2011kp,Jeon:2011cn,Jeon:2011vx,Jeon:2011sq,Hohm:2011ex,Hohm:2011zr,Hohm:2011dv,Hohm:2011cp,Hohm:2011nu,Hohm:2011si,ReidEdwards:2010vp,Coimbra:2011nw,Coimbra:2012yy,Albertsson:2008gq,Albertsson:2011ux,Copland:2011wx,Copland:2011yh,Kan:2011vg,Roiban:2012gi} for more applications of generalised geometry in string theory.} The worldvolume of the supermembrane theory as well as the worldsheet of string theory can also be rewritten in a manifestly duality invariant way by enlarging spacetime \cite{Duff:1990hn,Duff:1989tf,Tseytlin:1990nb,Tseytlin:1990va}. One can treat the metric and matter fields, as well as the spacetime and ``dual'' coordinates, on the same footing by constructing the supergravity as a non-linear realisation of the semi-direct product of $E_{11}$ with its first fundamental representation \cite{Berman:2011jh}.\footnote{The first fundamental representation gives rise to the spacetime and ``dual'' coordinates. The semi-direct product of $E_{11}$ with its first fundamental representation is its motion group, just as the Poincar\'{e} group is the motion group of the Lorentz group. \cite{Wigner:1939,Mackey:1951,Mackey:1952,Mackey:1955theory,Mackey:2004mathematical}} In general, a solution will depend on spacetime as well as dual coordinates, with the precise dependence satisfying a duality invariant sectioning condition \cite{Berman:2011cg}. The conventional spacetime solutions are then those which are independent of the dual coordinates. As will be studied in detail in section \ref{SDualities}, dualities will, in general, mix the conventional spacetime coordinates with dual coordinates. However, dualising only along isometries keeps the solutions independent of dual coordinates. This is the reason why the dualities emerge in the conventional picture only upon dimensional reduction. We stress, though, that the generalised geometry formulation allows dualities along any directions, even those which are not isometries.

In this paper we study the U-duality symmetries of the 11-dimensional supergravity arising as the low-energy theory of M-theory. In order to avoid difficulties with dualities along time-like directions \cite{Hull:1998vg,Hull:1998ym}, we Wick-rotate to obtain Euclidean solutions. We will restrict the dualities to only act in $d$ Euclidean directions, forming the group $E_d$\footnote{For $d<7$ we define $E_d$ to be the U-duality group acting in d Euclidean actions as given in table \ref{TDualityGroups}. Quantum effects will break the continuous symmetry group down into its discrete subgroup $E_d(Z)$ \cite{Hull:1994ys}.} \cite{Julia:1981sssg,ThierryMieg:1981sssg}, as given in table \ref{TDualityGroups}. We take the view that this restriction is artificial, solely for ease of calculation: U-duality should be allowed to act in all 11 Euclidean directions, even if they are not isometries, thus forming the proposed underlying symmetry group of the supergravity, $E_{11}$ \cite{Nicolai:1986jk,Duff:1985bv,deWit:1986mz,deWit:2000wu,West:2010rv,West:2001as,West:2003fc,West:2004kb,Kleinschmidt:2003jf,West:2004iz,West:2011mm,Riccioni:2007ni,Nicolai:2010zz,Damour:2002cu,Hillmann:2009ci}. We assume throughout that the metric is factorisable $g_{11} = g_{d} \oplus g_{\bar{d}}$, where $\bar{d} = 11-d$, and that the 3-form, $C_3$, and 6-form, $C_6$, potentials have non-zero components only along dualisable directions. Then for $d<6$ we can only have a non-vanishing 3-form potential in the dualisable directions. The case of $d = 3, 4$ is particularly simple because one only needs to consider winding modes of the M2-brane unlike the $d=5$ case where winding modes of the M5-brane will become important. We will consider higher dimensional duality groups in future publications.

\begin{table}
\centering
\begin{tabular}{|c|| c| c|}
\hline
d & $E_d$ & $H_d$ \\
\hline
3 & $SL(3) \otimes SL(2)$ & $SO(3) \otimes SO(2)$ \\
4 & $SL(5)$ & $SO(5)$ \\
5 & $SO(5,5)$ & $SO(5) \otimes SO(5)$ \\
6 & $E_6$ & $USp(8)$ \\
7 & $E_7$ & $SU(8)$ \\
8 & $E_8$ & $SO(16)$ \\
\hline
\end{tabular}
\caption{The U-duality groups $E_d$ and their maximal compact subgroups $H_d$}
\label{TDualityGroups}
\end{table}

We will first review the $E_3$ and $E_4$ algebras and how to construct the generalised metric as a non-linear realisation in section \ref{SAGenMetric}. These generalised metrics will unify the bosonic fields of 11-dimensional supergravity in a U-duality tensor which is used to write the low-energy effective action in a manifestly U-duality invariant way. Then in section \ref{SOmega} we study the local symmetries of the U-duality group which form $H_d$, the maximal compact subgroup of $E_d$, see table \ref{TDualityGroups}. We show how U-dualities transform the metric $g$ and 3-form potential, $C_3$, in section \ref{SDualities}, and compare the transformations to T-dualities of the 10-dimensional Euclidean supergravity in section \ref{STduality}. Finally, section \ref{SM2Brane} will consider the example of how M2-branes transform under U-duality.

\section{$E_{d}$ algebra and generalised metric} \label{SAGenMetric}
We begin by reviewing the U-duality algebra $E_{d}.$\footnote{We remind the reader that we follow the convention that $E_{d}$ always denotes the U-duality groups and their associated algebras as listed in table \ref{TDualityGroups}, even for $d \leq 5$.} In particular we describe how it arises by decomposing the $E_{11}$ algebra and heuristically construct the generalised metric as a non-linear realisation, as has been done in \cite{Berman:2011jh}. There the eleven dimensional duality algebra $E_{11}$ is decomposed into $E_d \otimes GL(\bar{d})$, with $\bar{d} = 11-d$, by deleting a node of the $E_{11}$ Dynkin diagram. The $E_d$ subalgebra acts in $d$ directions that can be dualised, while $GL(\bar{d})$ acts on the space transverse to the d directions. In \cite{Berman:2011jh} it was assumed that the transverse space is flat. We will not make this assumption here and we will see in section \ref{SDualities} that even if the transverse space is flat, in general, it will not be flat after the action of a U-duality.

\subsection{$d=3$ generalised metric} \label{S3Algebra}
We consider the $E_{11}$ algebra and decompose it into the duality algebra acting in 3 dimensions. This leaves the algebra of $\left(SL(2) \otimes SL(3)\right) \otimes GL(8)$. The $GL(8)$ is the rigid diffeomorphism group of the eight undualised directions and its non-linear realisation will give rise to gravity in those eight directions \cite{Isham:1971dv,Borisov:1974bn}, while $SL(2) \otimes SL(3) \equiv E_{3}$ is the U-duality group of three dimensions and its non-linear realisation will lead to the 11-dimensional supergravity fields. We decompose the $E_{11}$ generators according to this $3 + 8$ split and keep the level zero generators\footnote{The level zero generators are here the generators without mixed indices such as $K^i_{\ph{i}A}$, $R^{ijA}$ or $R_{ijA}$.}
\begin{equation}
 K^i_{\ph{i}j}, R^{123}, R_{123} ~\textrm{and}~ K^A_{\ph{A}B}
\end{equation}
where $i,j = 1, 2, 3$ and $A, B = 4, \ldots 11$. $K^i_{\ph{i}j}$ generate the $GL(3)$ which together with the totally antisymmetric generators $R^{123}$ and $R_{123}$ generate $E_{3}$ while $K^A_{\ph{A}B}$ generate $GL(8)$.

These generators satisfy
\begin{align}
 \left[ K^i_{\ph{i}j}, K^k_{\ph{k}l} \right] &= \delta^k_j K^i_{\ph{i}l} - \delta^i_l K^k_{\ph{k}j}, \label{EAlgebra1},\\
 \left[ K^A_{\ph{A}B}, K^C_{\ph{C}D} \right] &= \delta^C_B K^A_{\ph{A}D} - \delta^A_D K^C_{\ph{D}B} \label{EAlgebra2}, \\
 \left[ K^i_{\ph{i}j}, R^{klm} \right] &= 3 \delta^{[k}_j R^{|i|lm]}, \label{EAlgebra3}\\
 \left[ K^i_{\ph{i}j}, R_{klm} \right] &= - 3 \delta^{i}_{[k} R_{|j|lm]}, \label{EAlgebra4}\\
 \left[ R^{ijk}, R_{lmn} \right] = 18 \delta^{[ij}_{[lm} K^{k]}_{\ph{j}n]} &- 2 \delta^{ijk}_{lmn} \left( K^p_{\ph{p}p} + K^A_{\ph{A}A} \right), \label{EAlgebra5}
\end{align}
with all other commutators vanishing. The fully antisymmetrised Kronecker delta is defined as
\begin{equation}
 \begin{split}
 \delta^{ijk}_{lmn} &= \delta^{[i}_{[l} \delta^{j}_{m} \delta^{k]}_{n]} \\
 &= \frac{1}{3!} \left( \delta^i_l \delta^j_m \delta^k_n + \delta^k_l \delta^i_m \delta^j_n + \delta^j_l \delta^k_m \delta^i_n - \delta^i_l \delta^k_m \delta^j_n - \delta^k_l \delta^j_m \delta^i_n - \delta^j_l \delta^i_m \delta^k_n \right).
 \end{split}
\end{equation}

In three dimensions the relations \eqref{EAlgebra3} -- \eqref{EAlgebra5} simplify to
\begin{align}
 \begin{split}
 \left[ K^i_{\ph{i}j}, R^{123} \right] &= \delta^i_j R^{123}, \\
 \left[ K^i_{\ph{i}j}, R_{123} \right] &= - \delta^i_j R_{123}, \\
 \left[ R^{123}, R_{123} \right] = K^j_{\ph{j}j} &- \frac{1}{3} \left( K^j_{\ph{j}j} + K^A_{\ph{A}A} \right).
 \end{split}
\end{align}
In order to see the $SL(2) \otimes SL(3)$ structure explicitly, we define $K = K^i_{\ph{i}i}, \tilde{K}^i_{\ph{i}j} = K^i_{\ph{i}j} - \frac{1}{3} K \delta^i_j$ and $\bar{K} = \frac{2}{3} K - \frac{1}{3} K^A_{\ph{A}A}$. Then $SL(3)$ is generated by the trace-free generators $\tilde{K}^i_{\ph{i}j}$ while the $SL(2)$ is generated by $\bar{K}, R_{123}$ and $R^{123}$
\begin{equation}
 \begin{split}
 \begin{array}{ccc}
  \left[ \tilde{K}^i_{\ph{i}j}, \tilde{K}^k_{\ph{k}l} \right] = \delta^k_j \tilde{K}^i_{\ph{i}l} - \delta^i_l \tilde{K}^k_{\ph{k}j}, & \left[ \tilde{K}^i_{\ph{i}j}, R^{123} \right] = 0, & \left[ \tilde{K}^i_{\ph{i}j}, R_{123} \right] = 0, \\
  \left[ K, \tilde{K}^i_{\ph{i}j} \right] = 0, & \left[ \bar{K}, R_{123} \right] = -2 R_{123}, & \left[ \bar{K}, R^{123} \right] = 2 R^{123}, \end{array} \\
 \begin{array}{cc}
  \left[ R^{123}, R_{123} \right] = \bar{K}, & \left[ K^A_{\ph{A}B}, K^C_{\ph{C}D} \right] = \delta^C_B K^A_{\ph{A}D} - \delta^A_D K^C_{\ph{D}B} \end{array}.
 \end{split}
\end{equation}

To obtain the commutators of brane charges and spacetime momenta with the $E_d \otimes GL(\bar{d})$ algebra generators one considers the algebra of its motion group.\footnote{See \cite{Berman:2011jh} and references therein for details on how to construct the motion group.} We quote here the results found in \cite{Berman:2011jh}.
\begin{equation} \label{ECommGenCoord}
 \begin{split}
  \left[ K^i_{\ph{i}j}, P_l \right] &= -\delta^i_l P_j + \frac{1}{2} \delta^i_j P_l, \\
  \left[ K^i_{\ph{i}j}, Z^{kl} \right] &= 2 \delta^{[k}_j Z^{|i|l]} + \frac{1}{2} \delta^i_j Z^{kl}, \\
  \left[ R^{ijk}, P_l \right] &= 3 \delta^{[i}_l Z^{jk]}, \\
  \left[ R^{ijk}, Z^{kl} \right] &= 0, \\
  \left[ R_{ijk}, P_l \right] &= 0, \\
  \left[ R_{ijk}, Z^{lm} \right] &= 6 \delta^{lm}_{[ij} P_{k]}, \\
  \left[ K^A_{\ph{A}B}, P_C \right] &= - \delta^A_C P_B + \frac{1}{2} \delta^A_B P_C, \\
  \left[ K^A_{\ph{A}B}, P_i \right] &= \frac{1}{2} \delta^A_B P_i, \\
  \left[ K^A_{\ph{A}B}, Z^{kl} \right] &= \frac{1}{2} \delta^A_B Z^{kl}, \\
  \left[ K^i_{\ph{i}j}, P_A \right] &= \frac{1}{2} \delta^i_j P_A
 \end{split}
\end{equation}
and all others vanishing.

We want the generalised metric to act on the generalised coordinates
\begin{equation}
 X^M = \left( \begin{array}{c}
              x^i \\
              \frac{1}{\sqrt{2}} y_{ij} \\
              x^A
              \end{array} \right)
\end{equation}
which form an element of the algebra of the motion group
\begin{equation}
 \begin{split}
  X &= X^M T_M \\
  &= x^i P_i + x^A P_A + \frac{1}{2} y_{ij} Z^{ij},
 \end{split}
\end{equation}
where $T_M$ are the generators of the generalised translation group.

With each generator $K^i_{\ph{i}j}, R_{123}, R^{123}$ and $K^A_{\ph{A}B}$ we associate a field, $g_{ij}, \Omega^{123}, C_{123}$ and $g_{AB}$ respectively.\footnote{Recall that we assume that the metric is factorisable, i.e. $g_{iA} = 0$.} $C_{123}$ is just the 3-form potential of 11-dimensional supergravity while $\Omega^{123}$ is a trivector field which can be and usually is gauged away as we will explain in section \ref{SOmega}.\footnote{We expect the trivector field to play a role in non-geometric backgrounds where it may not be possible to gauge it away. This is similar to the role the bivector plays in string theory, see \cite{Grana:2008yw,Andriot:2011uh,Andriot:2012wx,Andriot:2012an,Aldazabal:2011nj,Dibitetto:2012rk}.} $g_{ij}$ are the components of the metric along the dualisable directions $x^i, i = 1, 2, 3$ and $g_{AB}$ the components along the other eight. We want to represent the $E_d$ generators by linear operators acting on the generalised coordinate basis $\left\{\frac{\partial}{\partial x^i}, \frac{\partial}{\partial y_{ij}}, \frac{\partial}{\partial x^A}\right\}$ which maps to the generalised translation generators in the motion algebra, $\left\{P_i, Z^{ij}, P_A\right\}$. Thus we consider the $E_d$ generators in the adjoint representation acting on an element of the generalised translation algebra $X = X^M T_M$. For example, we can find the appropriate linear operator corresponding to $-h^i_{\ph{i}j} K^j_{\ph{j}i}$ by finding its commutator with the generalised translation generators $P_i, Z^{ij}, P_A$,
\begin{align}
 \begin{split}
  \left[-h^i_{\ph{i}j} K^j_{\ph{j}i},P_l\right] &= -h^i_{\ph{i}j} \left( - \delta^j_l P_i + \frac{1}{2} \delta^j_i P_l \right) \\
  &= h^i_{\ph{i}l} P_i - \frac{1}{2} \left(\Tr h\right) P_l,
 \end{split} \\
 \begin{split}
  \left[-h^i_{\ph{i}j} K^j_{\ph{j}i},Z^{kl}\right] &= -h^i_{\ph{i}j} \left( 2\delta^{[k}_i Z^{|j|l]} + \frac{1}{2} \delta^j_i Z^{kl} \right) \\
  &= - h^k_{\ph{k}j} Z^{jl} + h^l_{\ph{l}j} Z^{jk} - \frac{1}{2} \left(\Tr h\right) Z^{kl},
 \end{split} \\
  \left[ -h^i_{\ph{i}j} K^j_{\ph{j}i},P_A\right] &= - \frac{1}{2} \left(\Tr h\right) P_A,
\end{align}
so that an element of the generalised translation algebra transforms as
\begin{equation}
 \begin{split}
 X &\rightarrow X' = \left(h^{i}_{\ph{i}j} - \frac{1}{2} \left(\Tr h\right) \delta^i_j\right)x^j P_i - \frac{1}{2} \left( \Tr h \right) \delta^A_B x^B P_A \\
 & + \left( -h^m_{\ph{m}k} \delta^n_l + h^n_{\ph{n}k} \delta^m_l - \frac{1}{2} \left(\Tr h\right) \delta^{mn}_{kl} \right) \frac{1}{2} y_{mn} Z^{kl}.
 \end{split}
\end{equation}
under the action of the adjoint operator corresponding to $-h^i_{\ph{i}j} K^j_{\ph{j}i}$. The linear operator acting on the generalised coordinate components is thus
\begin{equation}
 -h^i_{\ph{i}j} K^j_{\ph{j}i} = -\frac{1}{2} \Tr h \left( \begin{array}{ccc} h^i_{\ph{i}j} & 0 & 0 \\ 0 & - 2 h^{[m}_{\ph{[m}[k} \delta^{n]}_{\ph{n]}l]} & 0 \\ 0 & 0 & \delta^A_B \end{array} \right).
\end{equation}

Similarly, we find for the other $E_d$ generators
\begin{align}
 \frac{1}{3!}\Omega^{ijk} R_{ijk} &= \left( \begin{array}{ccc} 0 & \frac{1}{\sqrt{2}} \Omega^{imn} & 0 \\ 0 & 0 & 0 \\ 0 & 0 & 0 \end{array} \right), \\
 \frac{1}{3!}C_{ijk} R^{ijk} &= \left( \begin{array}{ccc} 0 & 0 & 0 \\ \frac{1}{\sqrt{2}} C_{klj} & 0 & 0 \\ 0 & 0 & 0 \end{array} \right), \\
 -k^A_{\ph{A}B} K^B_{\ph{B}A} &= -\frac{1}{2} \Tr k \left( \begin{array}{ccc} 1 & 0 & 0 \\ 0 & 1 & 0 \\ 0 & 0 & k^A_{\ph{A}B} \end{array} \right).
\end{align}

Using the exponential map we obtain the $E_{d} \otimes GL(\bar{d})$ group elements
\begin{align}
 \exp (-h^i_{\ph{i}j} K^j_{\ph{j}i}) &= |e^{h}|^{-1/2} \left( \begin{array}{ccc} \left( e^{h} \right)^i_{\ph{i}j} & 0 & 0 \\ 0 & \left( e^{-h} \right)^{[k}_{\ph{[k}[m} \left( e^{-h} \right)^{l]}_{\ph{l]}n]} & 0 \\ 0 & 0 & 1 \end{array} \right), \label{Eg}\\
 \exp(\frac{1}{3!}\Omega^{ijk} R_{ijk}) &= \left( \begin{array}{ccc} 1 & \frac{1}{\sqrt{2}} \Omega^{imn} & 0 \\ 0 & 1 & 0 \\ 0 & 0 & 1 \end{array} \right), \label{EO}\\
 \exp(\frac{1}{3!}C_{ijk} R^{ijk}) &= \left( \begin{array}{ccc} 1 & 0 & 0 \\ \frac{1}{\sqrt{2}} C_{klj} & 1 & 0 \\ 0 & 0 & 1 \end{array} \right), \label{EC}\\
 \exp (-k^A_{\ph{A}B} K^B_{\ph{B}A}) &= |e^{k}|^{-1/2} \left( \begin{array}{ccc} 1 & 0 & 0 \\ 0 & 1 & 0 \\ 0 & 0 & e^k \end{array} \right), \label{Et}
\end{align}
where $|e^h|$ denotes the determinant of the matrix $e^h$ and $1$ denotes the appropriate unit matrices, $\delta^i_j$, $\delta^{ij}_{kl}$ or $\delta^A_B$. We interpret these transformations as turning on a gravitational field in the dualisable directions (i.e. the metric $g_{ij}$), a 3-form field ($C_{ijk}$), trivector field ($\Omega^{ijk}$) and a gravitational field in the undualisable directions, respectively.

In flat space the generalised line element is just
\begin{equation}
dS^2 = \sum_i dx^i dx^i + \sum_A dx^A dx^A + \sum_{i,j} \frac{1}{2} dy_{ij} dy_{ij}
\end{equation}
and we can chose the generalised vielbein to be
\begin{equation}
L^{\bar{M}}_{\ph{\bar{M}}N} = \delta^{\bar{M}}_{\ph{\bar{M}}N},
\end{equation}
where $\bar{M}$ labels the flat tangent space generalised coordinates $X^{\bar{M}} = \left(x^{\bar{i}}, \frac{1}{\sqrt{2}} y_{\bar{i}\bar{j}}, x^{\bar{A}}\right)$. Then we can turn on a gravitational field and in the dualised directions a 3-form
\begin{equation}
 L \rightarrow \exp(\frac{1}{3!}C_{ijk} R^{ijk}) \exp (-h^i_{\ph{i}j} K^j_{\ph{j}i}) \exp (-h^A_{\ph{A}B} K^B_{\ph{B}A}) L.
\end{equation}
Upon identifying $e^h = \tilde{e}$ and $e^k = \tilde{e}_8$ as the metric vielbeins in the dualisable and undualisable directions and denoting by $|\tilde{e}|, |\tilde{e}_8|$ their respective determinants, we obtain the generalised vielbein
\begin{equation}
 L^{\bar{M}}_{\ph{M}N} = |\tilde{e} \tilde{e}_8|^{-1/2} \left( \begin{array}{ccc}
                        \tilde{e}^{\bar{j}}_{\ph{j}i} & 0 & 0 \\
                        \frac{1}{\sqrt{2}} C_{\bar{i}\bar{j}k} & e^{\ph{[\bar{k}}[i}_{[\bar{k}} e^{\ph{\bar{l}]}j]}_{\bar{l}]} & 0 \\
                        0 & 0 & \tilde{e}^{\bar{A}}_{\ph{A}B}
                        \end{array} \right).
\end{equation}
We will write this without explicit indices as
\begin{equation}
 L = |\tilde{e} \tilde{e}_8|^{-1/2} \left( \begin{array}{ccc}
                        \tilde{e} & 0 & 0 \\
                        \frac{1}{\sqrt{2}} e e C & e e & 0 \\
                        0 & 0 & \tilde{e}_8
                        \end{array} \right).
\end{equation}

We can also consider turning on a gravitational field and trivector in the dualisable directions to get
\begin{equation}
 L_\Omega = |\tilde{e}_{11}|^{-1/2} \left( \begin{array}{ccc}
                        \tilde{e} & \frac{1}{\sqrt{2}} \tilde{e} \Omega & 0 \\
                        0 & e e & 0 \\
                        0 & 0 & \tilde{e}_8
                        \end{array} \right),
\end{equation}
where $|\tilde{e}_{11}|$ is the determinant of the full 11-dimensional metric vielbein.

The two generalised vielbeins are related by a local $H \in H_3 \otimes SO(8)$ rotation, where $H_3 = SO(3) \otimes SO(2)$ and $H_3 \otimes SO(8)$ is a maximal compact subgroup of $E_3 \otimes GL(8)$,
\begin{equation}
 L_\Omega = H L.
\end{equation}
Because the two systems are related by an internal relation they are physically equivalent. This will be explored in more detail in section \ref{SOmega}. The generalised metrics of the gravity and 3-form system is
\begin{align}
 \gm_{MN} &= \left(L^T L\right)_{MN} \nonumber\\
 &= |g_{11}|^{-1/2} \left( \begin{array}{ccc}
           g_{ij} + \frac{1}{2} C_{i}^{\ph{i}pq} C_{pqj} & \frac{1}{\sqrt{2}} C_i^{\ph{i}mn}& 0 \\
           \frac{1}{\sqrt{2}} C^{kl}_{\ph{kl}j} & g^{k[m}g^{n]l} & 0 \\
           0 & 0 & g_{AB}
           \end{array} \right), \label{GM3}
\end{align}
where $|g_{11}|$ denotes the determinant of the 11-dimensional metric.

We will often drop the indices and write this as
\begin{align}
 \gm &= L^T L \\
 &= |g_{11}|^{-1/2} \left( \begin{array}{ccc}
           g + \frac{1}{2} C g^{-1} g^{-1} C & \frac{1}{\sqrt{2}} C g^{-1} g^{-1} & 0 \\
           \frac{1}{\sqrt{2}} g^{-1} g^{-1} C & g^{-1} g^{-1} & 0 \\
           0 & 0 & g_8
           \end{array} \right), \label{GM3}
\end{align}
where $g$ is the metric in the dualisable directions, i.e. has components $g_{ij}$, while $g_8$ is the metric in the eight transverse directions with components $g_{AB}$, and $g_{11}$ is the full 11-dimensional metric.

The other generalised metric with the trivector field is
\begin{align}
 \gm &= L_\Omega^T L_\Omega \\
 &= |g_{11}|^{-1/2} \left( \begin{array}{ccc}
           g & \frac{1}{\sqrt{2}} g \Omega & 0 \\
           \frac{1}{\sqrt{2}} \Omega g & g^{-1} g^{-1} + \frac{1}{2} \Omega g \Omega & 0 \\
           0 & 0 & g_8
           \end{array} \right).\footnotemark \label{GMO3}
\end{align} \footnotetext{The metrics $g_{11} = g_3 \otimes g_8$ in these two generalised metrics are different. The generalised metrics are equal though as they parametrise the same physical system using different variables. See section \ref{SOmega}.}

This generalised metric parametrises the coset $\frac{E_{3}\otimes GL(8)}{H_3 \otimes SO(8)}$ where $H_3 = SO(2) \otimes SO(3)$ is a maximal compact subgroup of $E_3 = SL(2) \otimes SL(3)$. $\gm$ transforms as a (0,2)-tensor under this ``extended'' U-duality such that for $U_e \in E_{3} \otimes GL(8)$,
\begin{equation}
 \gm \rightarrow U_e^T \gm U_e,
\end{equation}
or in terms of the generalised vielbein
\begin{equation}
 L \rightarrow L U_e,
\end{equation}
while it remains invariant under the action of $H_e \in H_3 \otimes SO(8)$\footnote{Recall that we are taking the full eleven dimensional theory to be Euclidean in order to avoid timlike dualities. Hence we use a $SO(8)$ local invariance of the $GL(8)$ rigid diffeomorphisms acting on the transverse space. For $E_d$ symmetry with $d<11$ one could also have chosen to keep a Lorentzian signature of the full 11-dimensional theory but demand that time always be part of the transverse undualisable space. In this case one would then have a local invariance of $H_3 \otimes SO(7,1)$.} since that just acts through an internal rotation of the generalised vielbein
\begin{equation}
 L \rightarrow H_e L.
\end{equation}
This is analogous to the way the conventional Lorentzian metric parameterises the coset $\frac{GL(d)}{SO(d-1,1)}$ where $GL(d)$ is the group of diffeomorphisms and $SO(d-1,1)$ is the local Lorentz symmetry\footnote{For Euclidean signature, as is used in this paper, $SO(d-1,1)$ is replaced by $SO(d)$.}, seen to act explicitly on the vielbein but not on the metric itself.

Because of the block-diagonal form, the $E_3$ and $GL(8)$ groups act independently, see equations \eqref{Eg} -- \eqref{Et}, so that a general element of the ``extended'' U-duality group $E_3 \otimes GL(8)$ can be written as
\begin{equation}
 U_e = \left( \begin{array}{cc}
              U & 0 \\
              0 & G \end{array} \right),
\end{equation}
where $U \in E_3$ is a $6\times6$ U-duality generator and $G \in GL(8)$.
\begin{equation}
 |g_{11}|^{-1/2} \left( \begin{array}{cc}
           g + \frac{1}{2} C g^{-1} g^{-1} C & \frac{1}{\sqrt{2}} C g^{-1} g^{-1} \\
           \frac{1}{\sqrt{2}} g^{-1} g^{-1} C & g^{-1} g^{-1} \\
           \end{array} \right)
\end{equation}
transforms as a (0,2)-tensor under a U-duality transformation $U \in E_3$ while
\begin{equation}
 |g_{11}|^{-1/2} g_{8}
\end{equation}
remains invariant under the U-dualities. This means that, in general, the transverse metric $g_{8}$ is not invariant under U-duality as we will see explicitly in section \ref{SDualities}. The $GL(8)$ group just acts as rigid diffeomorphisms on the transverse space such that for $G \in GL(8)$
\begin{equation}
 g_{8} \rightarrow G^T g_{8} G.
\end{equation}

Finally, it is important to stress that the conformal factor $|g_{11}|^{-1/2}$ in front of the generalised metric is crucial for it to transform as a tensor under U-duality. We will give the transformation rules explicitly in section \ref{SDualities} where it will become apparent that the conformal factor is necessary because of the transformation of the transverse space. One may also wish truncate the theory by dimensionally reducing along the transvserse eight-dimensional space. The generalised metric must then be constructed from the $E_3 = SL(2)\otimes SL(3)$ algebra alone. After a simple calculation one finds
\begin{equation}
 \gm_{MN} = |g|^{1/6} \left( \begin{array}{cc}
                     g_{ij} + \frac{1}{2} C_{i}^{\ph{i}pq} C_{pqj} & \frac{1}{\sqrt{2}} C_i^{\ph{i}mn} \\
                     \frac{1}{\sqrt{2}} C^{kl}_{\ph{kl}j} & g^{k[m}g^{n]l} \\
                     \end{array} \right).
\end{equation}
Because of the new conformal factor this generalised metric is non-dynamical, i.e. the Lagrangian of the truncated theory cannot be rewritten in terms of the generalised metric. This is analogous to the case of $SL(5)$ which was considered in \cite{Berman:2011jh}. Despite this drawback, this generalised metric can be expressed in terms of the trivector $\Omega^{ijk}$ instead of the 3-form $C_{ijk}$ in the same manners as in equation \eqref{GMO3}. This is important since $E_{11}$, and hence $E_3$, treat the 3-form and trivector on the same footing and happens only for the two conformal factors considered here. This, in fact, ensures that the generalised metric transforms as a tensor under U-duality. In conclusion, we see that the generalised metric transforms as a tensor under U-duality if the conformal factors $|g|^{1/6}$ or $|g_{11}|^{-1/2}$ are used for the truncated theory and 11-dimensional one, respectively. With a different conformal factor, the generalised metric will not transform correctly under dualities.

\subsection{$d = 4$ generalised metric}
For the duality group acting in four dimensions we decompose $E_{11}$ into $SL(5) \otimes GL(7)$, where $SL(5) = E_4$ is the U-duality group in four dimensions. The algebra of the motion group is again given by equations \eqref{EAlgebra1} -- \eqref{EAlgebra5} and \eqref{ECommGenCoord} and the generalised metric can be constructed in an analogous fashion to obtain
\begin{equation}
 \gm = |g_{11}|^{-1/2} \left( \begin{array}{ccc}
                                g + \frac{1}{2} C g^{-1} g^{-1} C & \frac{1}{\sqrt{2}} C g^{-1} g^{-1} & 0 \\
                                \frac{1}{\sqrt{2}} g^{-1} g^{-1} C & g^{-1} g^{-1} & 0 \\
                                0 & 0 & g_7 \end{array} \right) \label{GM4}
\end{equation}
and expressed in terms of the trivector $\Omega_3$ we get
\begin{equation}
 \gm_{\Omega} = |g_{11}|^{-1/2} \left( \begin{array}{ccc}
           g & \frac{1}{\sqrt{2}} g \Omega & 0 \\
           \frac{1}{\sqrt{2}} \Omega g & g^{-1} g^{-1} + \frac{1}{2} \Omega g \Omega & 0 \\
           0 & 0 & g_7
           \end{array} \right). \label{GMO4}
\end{equation}
A similar generalised metric was found in \cite{Duff:1990hn,Hull:2007zu} in the four-dimensional case. However, there the conformal factor was missed. While the resultant generalised metric still unifies the gauge symmetries of the bosonic fields --- the diffeomorphism symmetry and gauge symmetry of the 3-form, $C_3$ ---, it does not transform as a tensor under U-duality, as explained in the discussion on the three-dimensional generalised metric above.

Because the generalised metrics for $d=3$ and $d=4$ have a similar structure most considerations apply equally to both. Thus, the following comments will apply to both except where explicitly stated otherwise.

\subsection{Duality invariant action}
The action of the U-duality group $E_d$ has been ``geometrised'' in the sense that it can be seen to arise from rigid transformations of the generalised coordinates
\begin{equation}
 X^M = \left( \begin{array}{ccc}
              x^i \\
              \frac{1}{\sqrt{2}} y_{ij} \\
              x^A
              \end{array} \right).
\end{equation}
Under the extended U-duality group $U_e \in E_d \otimes GL(\bar{d})$, where $d + \bar{d} = 11$, the generalised coordinates transform as vectors
\begin{equation}
 X \rightarrow U_e^{-1} X.
\end{equation}
The generalised line element $dS^2 = \gm_{MN} dX^M dX^N$ then remains invariant under the extended U-duality transformations.

The 11-dimensional Lagrangian can be written in a manifestly invariant way by expressing it in terms of the generalised metric $\gm$ and generalised coordinates $X$. This requires an integration by parts so it becomes a first order Lagrangian\footnote{The Chern-Simons term vanishes because here the 3-form has non-vanishing components only in the dualisable directions.}
\begin{align}
 \mathcal{L} &= \sqrt{|g_{11}|} \left(R - \frac{1}{48} F^2\right) \nonumber \\
 \begin{split}
 &\sim \sqrt{|g_{11}|} \left[ \frac{1}{2} \partial_a \ln|g_{11}| \partial_b g^{ab} - \frac{1}{2} g^{ab} \partial_a g^{cd} \partial_c g_{bd} \right. \nonumber \\
 & \left. + \frac{1}{4} g^{ab} \partial_a \ln|g_{11}| \partial_b \ln|g_{11}| + \frac{1}{4} g^{ab} \Tr \left( \partial_a g_{11}^{-1} \partial_b g_{11} \right) - \frac{1}{48} F^2 \right],
 \end{split}
\end{align}
where $|g_{11}|$ denotes the determinant of the 11-dimensional metric and by $\sim$ we mean equal up to integration by parts. In addition to $\gm$ and $X$ we can also use the U-duality invariant $|g_{11}|^{-1/2} g_{\bar{d}}$ where $g_{\bar{d}}$ is the metric in the undualisable $\bar{d}$ directions. This term remains invariant under the action of the extended U-duality group $E_d \otimes GL(\bar{d})$ as long as its spacetime indices are contracted appropriately. We find that for $d = 4$ the Lagrangian can be written as
\begin{equation}
 \begin{split}
 \mathcal{L} &\sim \frac{1}{12} \gm^{MN} \partial_M \gm^{PQ} \partial_N \gm_{PQ} - \frac{1}{2} \gm^{MN} \partial_M \gm^{PQ} \partial_P \gm_{NQ} \\
 & \quad + \frac{1}{108} \gm^{MN} \left( \gm^{KL} \partial_M \gm_{KL} \right) \left( \gm^{PQ} \partial_N \gm_{PQ} \right) \\
 & \quad + \frac{1}{6} \gm^{MN} \partial_M \left( |g_{11}|^{1/2} g^{AB} \right) \partial_N \left( |g_{11}|^{-1/2} g_{AB} \right). \label{ELEEA}
 \end{split}
\end{equation}

The action written in terms of the generalised metric reduces to the usual Einstein-Hilbert-3-form action if we take $\partial_{y} = 0$, i.e. no dependence on the ``dual'' coordinates.\footnote{We stress that the Lagrangian for the truncated theory, i.e. where the theory is dimensionally reduced along the transverse space, cannot be rewritten in terms of its generalised metric because the appropriate generalised metric is non-dynamical as mentioned in section \ref{S3Algebra} and as discussed more explicitly for the four-dimensional case in \cite{Berman:2011jh}.} As mentioned before, the restriction of U-dualities to act only in $d = 3,4$ directions is artificial and the full duality group should be taken to act in all eleven directions. The restriction is simply used to make the calculation more tractable. When the full duality group, $E_{11},$\footnote{We use the term ``group'' to describe $E_{11}$ in the same way as \cite{West:2001as}.} is considered and the appropriate generalised metric, $\gm_{11}$, is constructed, we expect the Lagrangian to be expressible purely in terms of $\gm_{11}$.

For now, though, this action suffices and is manifestly duality invariant so that U-dualities turn solutions of the generalised field equations into new solutions. Because the coordinates transform as well under a U-duality, even if we start with a solution of the Einstein-3-form system, i.e. one satisfying the sectioning condition $\partial_{y} = 0$, the transformed solution may no longer preserve this condition. Thus, the transformed solution may contain dual coordinates, $y_{ij}$. However, if the duality transformations are taken along isometries then no dual coordinates will be introduced into the solution.

\section{$\Omega$-field and local symmetry} \label{SOmega}
The action of the $GL(\bar{d})$ group is nothing but rigid diffeomorphisms acting on the transverse undualisable space and we will thus ignore it. We therefore only consider the action of the U-duality group $E_d$, and restrict the generalised metric to the components along dualisable coordinates and their dual coordinates. The transverse coordinates $x^A \rightarrow x^A$ will always transform trivially under this group and we can, and will, simply ignore the coordinates $x^A$ except where mentioned explicitly.

As discussed in section \ref{SAGenMetric}, when we construct the generalised metric from its U-duality algebra, we find that it should contain three fields: the metric $g$, the 3-form potential $C_3$ and a tri-vector potential $\Omega_3$. However, the two matter fields $C_3$ and $\Omega_3$ are not independent but can be rotated into one another by a local symmetry $H \in H_d$. Thus, the true physical degrees of freedom can be described by either $\gm_C(g_{11}, C_3)$ or $\gm_\Omega(\og_{11}, \Omega_3)$, where the metrics $g_{11}$ and $\og_{11}$ are different in general. It is useful to have a way of rewriting a system with metric and 3-form into one with metric and tri-vector because, when it comes to dualities, one of the descriptions may prove more useful as I will explain in section \ref{SDualities}.

The generalised metric written with respect to $C_3$ is given in equations \eqref{GM3} and \eqref{GM4}. On the other hand, when constructed with a metric and tri-vector field it is
\begin{equation}
 \begin{split}
 \gm_\Omega &= |\og_{11}|^{-1/2} \left( \begin{array}{ccc}
                    \og_{ij} & \frac{1}{\sqrt{2}} \og_{ik} \Omega^{kmn} & 0 \\
                    \frac{1}{\sqrt{2}} \Omega^{klm} \og_{mj} & \og^{k[m} \og^{n]l} + \frac{1}{2} \Omega^{klp} \og_{pq} \Omega^{qmn} & 0 \\
                    0 & 0 & \og_{\bar{d}}
                    \end{array} \right) \\
 &\equiv |\og_{11}|^{-1/2} \left( \begin{array}{ccc}
                    \og_{ij} & \frac{1}{\sqrt{2}} \Omega_{i}^{\ph{i}mn} & 0 \\
                    \frac{1}{\sqrt{2}} \Omega^{kl}_{\ph{kl}j} & \og^{k[m} \og^{n]l} + \frac{1}{2} \Omega^{klp} \Omega_p^{\ph{p}mn} & 0 \\
                    0 & 0 & \og_{\bar{d}}
                    \end{array} \right).
 \end{split}
\end{equation}
We use the convention that indices on $\Omega_3$ are always lowered with $\og$ unless specified otherwise.

Because the generalised metrics $\gm_C$ and $\gm_\Omega$ are related by internal rotations they must be equal. From the equation
\begin{equation}
 \gm_C(g_{11},C_3) = \gm_\Omega(\og_{11},\Omega_3)
\end{equation}
we can read off the relations between the fields $g, C$ and $\og, \Omega$. In three dimensions we find that
\begin{equation} \label{3CtoOmega}
 \begin{split}
 \og_{ij} &= g_{ij} \left(1 + V^2\right)^{2/3}, \\
 \Omega^{ijk} &= \frac{\epsilon^{ijk}V}{1+V^2} = \frac{g^{im} g^{jn} g^{ko} C_{mno}}{1+V^2}, \\
 \og_{AB} &= g_{AB} \left(1 + V^2\right)^{-1/3},
 \end{split}
\end{equation}
where $V = \frac{1}{3!} \epsilon^{ijk} C_{ijk} $ and $\epsilon^{ijk}$ is the Levi-Civita tensor in the three dimensions to be dualised.

The inverse identities are
\begin{equation} \label{3OmegatoC}
 \begin{split}
 g_{ij} &= \og_{ij} \left(1+W^2\right)^{-2/3}, \\
 C_{ijk} &= \frac{\bar{\epsilon}_{ijk}W}{1+W^2} = \frac{\og_{im} \og_{jn} \og_{ko} \Omega^{mno}}{1+W^2}, \\
 g_{AB} &= \og_{AB} \left(1 + W^2\right)^{1/3},
 \end{split}
\end{equation}
where $W = \frac{1}{3!} \bar{\epsilon}_{ijk} \Omega^{ijk}$ and $\bar{\epsilon}_{ijk}$ is the Levi-Civita tensor with respect to $\og_3$.

In four dimensions we have
\begin{equation}
 \begin{split} \label{4CtoOmega}
 \og_{ij} &= g_{ik} \left( \left(1 + V^2\right) \delta^k_j - V^j V_k \right) \left(1+V^2\right)^{-1/3}, \\
 \Omega^{ijk} &=  \frac{\epsilon^{ijkl}V_l}{1+V^2} = \frac{g^{im} g^{jn} g^{ko} C_{mno}}{1+V^2}, \\
 \og_{AB} &= g_{AB} \left(1 + V^2\right)^{-1/3},
 \end{split}
\end{equation}
where $V^i = \frac{1}{3!} \epsilon^{ijkl} C_{jkl}$, $V^2 = V^i V_i$ and $\epsilon^{ijkl}$ is the Levi-Civita tensor with resepect to $g_4$ and inverse identities
\begin{equation} \label{4OmegatoC}
 \begin{split}
 g_{ij} &= \og_{ik} \left( \delta^k_j + W^k W_j \right) \left(1+W^2\right)^{-2/3}, \\
 C_{ijk} &= \frac{\bar{\epsilon}_{ijkl} W^l}{1+W^2} = \frac{\og_{im} \og_{jn} \og_{ko} \Omega^{mno}}{1+W^2}, \\
 g_{AB} &= \og_{AB} \left(1 + W^2\right)^{1/3},
 \end{split}
\end{equation}
where similarly $W_i = \frac{1}{3!} \bar{\epsilon}_{ijkl} \Omega^{jkl}$, $W^2 = W^i W_i$ and $\bar{\epsilon}_{ijkl}$ is the Levi-Civita tensor with respect to $\og_4$.

\section{U-duality} \label{SDualities}
We will now discuss the orbits of U-duality acting in three and four directions. The three-dimensional case will be explained in detail, but since the four-dimensional case is very similar, details which are the same will be omitted.

\subsection{Duality in d=3, $SL(2) \otimes SL(3)$}
The U-duality group acting in three directions $E_{3} = SL(2) \otimes SL(3)$ is generated by the following elements, given in the representation acting on the generalised metric,
\begin{align}
 U_{\omega} \equiv \exp(\frac{1}{3!}\omega^{ijk} R_{ijk}) &= \left( \begin{array}{cc} 1 & \frac{1}{\sqrt{2}} \omega \\ 0 & 1 \end{array} \right), \\
 U_c \equiv \exp(\frac{1}{3!}c_{ijk} R^{ijk}) &= \left( \begin{array}{cc} 1 & 0 \\ \frac{1}{\sqrt{2}} c & 1 \end{array} \right), \\
 U_{GL(3)} \equiv \exp(-h^i_{\ph{i}j} K^j_{\ph{j}i}) &=  |A|^{-1/2} \left( \begin{array}{cc} A & 0 \\ 0 & A^{-T} A^{-T} \end{array} \right),
\end{align}
where $A = e^h$ denotes a $3x3$ matrix.
Thus, the U-duality group acts through $GL(3)$ of rigid diffeomorphisms acting on the metric $g_3$, and constant shifts in the 3-form potential $C_3$ and tri-vector potential $\Omega_3$. These are all gauge transformations of the fields
\begin{align}
 g_3 & \rightarrow A^T g_3 A, \\
 C_3 & \rightarrow C_3 + c, \\
 \Omega_3 & \rightarrow \Omega_3 + \omega, \\
 g_8 & \rightarrow g_8,
\end{align}
but because of the non-linear relation between $\left(g_{11}, C_3\right)$ and $\left(\og_{11}, \Omega_3\right)$, a gauge transformation of $\Omega_3$ will relate two physically distinct systems expressed through a metric $g_{11}$ and 3-form $C_3$.

\subsubsection{SL(3) subgroup} \label{SSL3}
The $SL(3)$ subgroup acts by interchanging the three dualisable directions. This is a ``geometric'' $SL(3)$: if the three dualisable directions form a 3-torus, the $SL(3)$ is just its modular group. The group elements are obtained by exponentiating the traceless generators of the algebra $\tilde{K}^{i}_{\ph{i}j}$. Their representation thus consists of matrices of unit determinant
\begin{equation}
 U_{SL(3)} = \left( \begin{array}{cc}
                    A & 0 \\
                    0 & A^{-T} A^{-T}
                    \end{array} \right),
\end{equation}
or with indices made explicit
\begin{equation} \label{ESL3}
 \left(U_{SL(3)}\right)^M_{\ph{M}N} = \left( \begin{array}{cc}
                                A^i_{\ph{i}j} & 0 \\
                                0 & \left(A^{-1}\right)_{[m}^{\ph{[m}[k} \left(A^{-1}\right)_{n]}^{\ph{n]}l]}
                                \end{array} \right).
\end{equation}
Here $\det A= 1$ so $\det U = 1$. The transformation just interchanges the dualisable directions
\begin{equation}
 \begin{split}
 x^i &\rightarrow \left(A^{-1}\right)^i_{\ph{i}j} x^j, \\
 y_{ij} & \rightarrow A_{i}^{\ph{i}k} A_j^{\ph{j}l} y_{kl}, \\
 g_{ij} &\rightarrow A_{i}^{\ph{i}k} g_{kl} A^{l}_{\ph{l}j}.
 \end{split}
\end{equation}

Each pair of directions is acted on by a $SL(2)$ subgroup belonging to $SL(3)$. If we regard the two directions as forming a $T^2$ we get the usual K\"{a}hler parameter
\begin{equation}
 \tau = \frac{1}{g_{11}} \left( g_{12} + i \sqrt{|g_2|} \right),
\end{equation}
where we have taken the directions $x^1$ and $x^2$ as a pair and define $|g_2| = g_{11} g_{22} - g_{12}^2$. The action of the $SL(2)$, even if the directions do not form a $T^2$, is then the usual
\begin{equation}
 \tau \rightarrow \frac{a \tau + b}{c \tau + d},
\end{equation}
\begin{equation}
 ad - bc = 1 \nonumber.
\end{equation}
For example, $\tau \rightarrow a^2 \tau$, i.e. $a = 1/d$ and $b = c = 0$ above, is generated by
\begin{equation}
 A = \left( \begin{array}{ccc} a & 0 & 0 \\ 0 & \frac{1}{a} & 0 \\ 0 & 0 & 1 \end{array} \right),
\end{equation}
$\tau \rightarrow \tau + b$ is generated by
\begin{equation}
 A = \left( \begin{array}{ccc} 1 & b & 0 \\ 0 & 1 & 0 \\ 0 & 0 & 1 \end{array} \right),
\end{equation}
and $\tau \rightarrow \frac{\tau}{c\tau + 1}$ is generated by
\begin{equation}
 A = \left( \begin{array}{ccc} 1 & 0 & 0 \\ c & 1 & 0 \\ 0 & 0 & 1 \end{array}\right).
\end{equation}

\subsubsection{SL(2) subgroup} \label{SSL2}
The $SL(2)$ contains the non-trivial action of U-duality on the metric $g_{11}$ and 3-form $C_3$. The group is generated by the elements
\begin{align}
 U_{\omega} \equiv \exp(\omega^{123} R_{123}) &= \left( \begin{array}{cc} 1 & \frac{1}{\sqrt{2}} \omega \\ 0 & 1 \end{array} \right), \\
 U_c \equiv \exp(c_{123} R^{123}) &= \left( \begin{array}{cc} 1 & 0 \\ \frac{1}{\sqrt{2}} c & 1 \end{array} \right), \\
 U_{\alpha} \equiv \exp(h K) &= \left( \begin{array}{cc} \alpha & 0 \\ 0 & \alpha^7 \end{array} \right),
\end{align}
where $\alpha = e^{h/2}$ is a scalar.

$U_\alpha$ just generates scalings of the coordinates so that
\begin{equation}
 \begin{split}
 x &\rightarrow \alpha^{-1} x, \\
 y &\rightarrow \alpha^{-7} y, \\
 g_3 &\rightarrow \alpha^{-4} g_3, \\
 C_3 &\rightarrow \alpha^{-6} C_3, \\
 g_8 &\rightarrow g_8.
 \end{split}
\end{equation}

$U_c$ shifts the 3-form potential by a constant. But because generalised geometry geometrises the duality transformations, the generalised coordinates will transform as well
\begin{equation}
 \begin{split}
 x^i & \rightarrow x^i, \\
 y_{ij} & \rightarrow y_{ij} - c_{ijk} x^k, \\
 g_{ab} & \rightarrow g_{ab}, \\
 C_{ijk} & \rightarrow C_{ijk} + c_{ijk}.
 \end{split}
\end{equation}

On the other hand, $U_\omega$ shifts the trivector potential $\Omega$ by a constant. The transformation is trivial written in terms of $\og_{11}, \Omega_3$
\begin{equation}
 \begin{split}
 x^i & \rightarrow x^i - \frac{1}{2} \omega^{ijk} y_{jk}, \\
 y_{ij} & \rightarrow y_{ij}, \\
 \og_{ab} & \rightarrow \og_{ab}, \\
 \Omega^{ijk} & \rightarrow \Omega^{ijk} + \omega^{ijk}.
 \end{split}
\end{equation}
All three transformations look like gauge transformations in a certain frame (i.e. $g_{11}, C_3$ or $\og_{11}, \Omega_3$) and thus describe ``equivalent'' systems as seen from the level of M-branes. The $C$-shift is a gauge transformation as seen in the $\left(g_{11}, C_3\right)$ frame but is a non-trivial transformation in the $\left(\og_{11}, \Omega_3\right)$ frame. Similarly, the $\Omega$-shift is a gauge transformation in the $\left(\og_{11}, \Omega_3\right)$ frame but gives a non-trivial transformation in the $\left(g_{11}, C_3\right)$ frame.

Using the relations \eqref{3CtoOmega} and \eqref{3OmegatoC} we can calculate the action of $U_\Omega$ as seen in the $\left(g_{11}, C_3\right)$ frame. For $\Omega^{123} = A$ we find
\begin{equation}
 \begin{split} \label{3OmegaDuality}
 g'_{ij} &= g_{ij} \left((1+AC_{123})^2 + A^2 |g_3| \right)^{-2/3}, \\
 g'_{AB} &= g_{AB} \left((1+AC_{123})^2 + A^2 |g_3| \right)^{1/3}, \\
 C'_{123} &= \frac{C_{123} \left(1 + A C_{123}\right) + A|g_3|}{\left(1+AC_{123}\right)^2 + A^2 |g_3|}.
 \end{split}
\end{equation}
This kind of transformation is related to the Lunin-Maldacena deformation \cite{Lunin:2005jy,CatalOzer:2005mr,Frolov:2005dj} of supergravities that was found by studying their holographic duals.

The transformations $U_c, U_\omega, U_\alpha$ together generate a $SL(2)$ subgroup that can be seen to act on the K\"{a}hler parameter
\begin{equation}
 E = i \sqrt{|g_3|} + C_{123},
\end{equation}
or in covariant notation
\begin{equation}
 E = \sqrt{|g_3|} \left(i + \star C_3 \right),
\end{equation}
where $\star$ is the Hodge dual operator in 3-dimensions $\star C_3 \equiv \frac{1}{3!} \epsilon^{ijk} C_{ijk}$ and $\epsilon^{123} = \frac{1}{\sqrt{g_3}}$ is the 3-d Levi-Civita tensor. Under this $SL(2)$ group $E$ transforms as
\begin{equation}
 E \rightarrow \frac{a E + b}{c E + d}. \label{EKaehler}
\end{equation}
For a M2-brane this corresponds to the K\"{a}hler parameter for T-duality transformations, as found in \cite{Sen:1995cf}.

We find that the duality transformations act through
\begin{align}
 \textrm{dilatations, } U_\alpha: E & \rightarrow \alpha^{-6} E \label{EKaehlerA}, \\
 C_3 \textrm{ shifts, } U_c: E & \rightarrow E + c_{123} \label{EKaehlerC}, \\
 \textrm{and } \Omega_3 \textrm{ shifts, } U_\omega: E & \rightarrow \frac{E}{1+\omega^{123} E}. \label{EKaehlerO}
\end{align}

There is also a component which generates what we will call a ``Buscher duality''. We will show that it reduces to a pair of Buscher dualities \cite{Buscher:1987qj,Buscher:1987sk} of the Type IIA background in section \ref{STduality}. It acts on the K\"{a}hler parameter as
\begin{equation}
 E \rightarrow - \frac{1}{E},
\end{equation}
and can be constructed from three successive $SL(2)$ transformations
\begin{align}
 U_B &= \left( \begin{array}{cc}
            1 & 0 \\
            \frac{1}{\sqrt{2}} C & 1
            \end{array} \right)
     \left( \begin{array}{cc}
            1 & \frac{1}{\sqrt{2}} \Omega \\
            0 & 1
            \end{array} \right)
     \left( \begin{array}{cc}
            1 & 0 \\
            \frac{1}{\sqrt{2}} C & 1
            \end{array} \right) \nonumber \\
 &= \left( \begin{array}{cc}
            0 & \frac{1}{\sqrt{2}} \Omega \\
            \frac{1}{\sqrt{2}} C & 0
            \end{array} \right), \label{EBuscherGenerated}
\end{align}
where $\frac{1}{2} \Omega C = -1$, i.e.
\begin{equation}
 \begin{split}
 \Omega^{123} &= A, \\
 C_{123} &= - 1/A.
 \end{split}
\end{equation}

The transformation of the fields in the $\left(g_{11}, C_3\right)$ frame can then seen to be
\begin{equation}
 \begin{split} \label{3BuscherDuality}
  g'_{ij} &= g_{ij} \left( A^2 \left( C_{123}^2 + |g_3| \right) \right)^{-2/3}, \\
  g'_{AB} &= g_{AB} \left( A^2 \left( C_{123}^2 + |g_3| \right) \right)^{1/3}, \\
  C'_{123} &= - \frac{C_{123}}{A^2 \left( C_{123}^2 + |g_3| \right)}.
 \end{split}
\end{equation}
For a 3-torus that is dualised, i.e. $g_{11} = R_1^2, g_{22} = R_2^2, g_{33} = R_3^2$ and $C_{123} = 0$ we find with $A = 1$
\begin{equation} \label{EMBuscher}
 \begin{split}
  R_1 & \rightarrow R_1^{1/3} R_2^{-2/3} R_3^{-2/3}, \\
  R_2 & \rightarrow R_1^{-2/3} R_2^{1/3} R_3^{-2/3}, \\
  R_3 & \rightarrow R_1^{-2/3} R_2^{-2/3} R_3^{1/3}.
 \end{split}
\end{equation}
This actually reduces to a pair of Buscher dualities of string theory as we will show in section \ref{STduality}.

Finally, for sake of completeness, we will define a K\"{a}hler parameter for the $\og, \Omega$ system
\begin{equation}
 \begin{split}
 \bar{E} &= \frac{i}{\sqrt{|\og_3|}} + \Omega^{123} \\
 &= \frac{1}{\sqrt{|\og_3|}} \left(i + W\right).
 \end{split}
\end{equation}

This K\"{a}hler parameter now transforms as
\begin{equation} \label{OE11Kaehler}
 \bar{E} \rightarrow \frac{a\bar{E} + b}{c\bar{E} + d},
\end{equation}
where $ad - bc = 1$ and we find
\begin{align}
 U_\alpha: \bar{E} & \rightarrow \alpha^{6} \bar{E} \label{OE11KaehlerA}, \\
 U_\omega: \bar{E} & \rightarrow \bar{E} + \omega^{123} \label{OE11KaehlerO}, \\
 U_c: \bar{E} & \rightarrow \frac{\bar{E}}{1+c_{123} \bar{E}} \label{OE11KaehlerC}.
\end{align}

\subsection{Duality in d=4, $SL(5)$}

For $d=4$ the U-duality group $SL(5)$ similarly contains a $SL(4)$ subgroup generating rigid diffeomorphisms. This $SL(4)$ subgroup is once again generated by exponentiating the traceless algebra generators $\tilde{K}^{i}_{\ph{i}j}$. Its elements are thus of the form

\begin{equation}
 U_{SL(4)} = \left( \begin{array}{cc}
                    A & 0 \\
                    0 & A^{-T} A^{-T}
                    \end{array} \right),
\end{equation}
wwhere $\det A = 1$ and $A^{-T} A^{-T}$ again represents the antisymmetrised product as has been made explicit in equation \eqref{ESL3}. Once again these transformations only interchange the dualisable coordinates amongst themselves.

For every set of three spacetime directions, the quotient group $SL(5)/SL(4)$ contains a $SL(2)$ subgroup
\begin{equation}
 \begin{split}
 SL(2)_{123} &= \left\{ U_{\alpha}, U_{\omega^{123}}, U_{c_{123}} \right\}, \\
 SL(2)_{124} &= \left\{ U_{\alpha}, U_{\omega^{124}}, U_{c_{124}} \right\}, \\
 SL(2)_{134} &= \left\{ U_{\alpha}, U_{\omega^{134}}, U_{c_{134}} \right\}, \\
 SL(2)_{234} &= \left\{ U_{\alpha}, U_{\omega^{234}}, U_{c_{234}} \right\},
 \end{split}
\end{equation}
although these do not commute amongst themselves. For each of these $\frac{d!}{(d-3)!3!}$ $SL(2)$ subgroups we can define a K\"{a}hler parameter just as before, e.g.
\begin{align}
 E_{123} &= \sqrt{|g_4|} \left(i g^{44} + \left(\star C_3\right)^4 \right) \\
 &= i \sqrt{|g_3|} + C_{123},
\end{align}
where $|g_3| = |g_4| g^{44}$. Under $SL(2)_{123}$ this K\"{a}hler parameter transforms as in equation \eqref{EKaehler}. The $C$-shifts and dilatations act exactly as before ,whereas there is a difference when considering $\Omega$-shifts. For these we find that the transformations are the same for $g_{\alpha\beta}$ and $C_{\alpha\beta\gamma}$ as in the 3-d case, where $\alpha, \beta, \gamma = 1, 2, 3$. However, the mixed and transverse components transform differently if there are non-zero mixed components of the metric, $g_{\alpha 4}$, and 3-form, $C_{\alpha\beta 4}$. Explicitly, we have for $\Omega^{123} = A$
\begin{equation}
 \begin{split}
 g'_{ij} &= \left(g_{ij} + A\sqrt{|g_4|} V_i \delta^4_j + A\sqrt{|g_4|} V_j \delta^4_i + A^2 |g_4| \left(1+V^2\right) \delta^4_i \delta^4_j \right) \\
 & \quad \times \left( \left(1 + AC_{123}\right)^2 + A^2 |g_4| g^{44} \right)^{-2/3}, \\
 g'_{AB} &= g_{AB} \left((1+AC_{123})^2 + A^2 |g_4| g^{44} \right)^{1/3}, \\
 C'_{ijk} &= \frac{C_{ijk} \left(1+AC_{123}\right)+A \sqrt{|g_4|} \epsilon_{ijkl}g^{l4}}{\left(1+AC_{123}\right)^2+A^2|g_4| g^{44}},
 \end{split}
\end{equation}
and letting $|g_{\alpha\beta}| = |g_3|$, we can write this as
\begin{equation}
 \begin{split} \label{E4OmegaDuality}
 g'_{\alpha\beta} &= g_{\alpha\beta} \left( \left(1 + A C_{123} \right)^2 + A^2 |g_3| \right)^{-2/3}, \\
 g'_{\alpha4} &= \left(g_{\alpha4} - \frac{1}{3!} A g_{\alpha i} \sqrt{|g_4|} \epsilon^{ijkl} C_{jkl} \right) \left( \left(1 + A C_{123} \right)^2 + A^2 |g_3| \right)^{-2/3}, \\
 g'_{44} &= g_{44} \left( \left(1 + AC_{123}\right)^2 + A^2 |g_3| \right)^{1/3} + \Big(A^2 |g_3| \left(\frac{1}{g^{44}} - g_{44}\right) \Big. \\
 & \Big. \quad + A^2 |g_4| g_{4\alpha} V^4 V^\alpha + A^2 |g_4| V^{\alpha} V_{\alpha} + 2 A g_{4\alpha} V^\alpha \sqrt{|g_4|} \Big) \left( \left(1 + A C_{123} \right)^2 + A^2 |g_3| \right)^{-2/3}, \\
 g'_{AB} &= g_{AB} \left((1+AC_{123})^2 + A^2 |g_3| \right)^{1/3}, \\
 C'_{ijk} &= \frac{C_{ijk} \left(1+AC_{123}\right)+A \sqrt{|g_4|} \epsilon_{ijkl}g^{l4}}{\left(1+AC_{123}\right)^2+A^2 |g_3|}.
 \end{split}
\end{equation}

\section{Type IIA T-duality} \label{STduality}
11-dimensional supergravity compactified on $S^1$ of zero radius gives rise to the IIA ten-dimensional supergravity. Similarly, the generalised geometry formulations are related. In \cite{West:2010ev} the non-linear realisation of $E_{11}$ is used to obtain the generalised metric of the type IIA supergravity while \cite{Thompson:2011uw} directly reduces the $E_4 = SL(5)$ generalised metric to obtain the $SO(3,3)$ generalised metric of string theory. Here we discuss the connection between the U-duality transformations to the T-duality symmetry of the ten-dimensional supergravity. Here T-duality always means those dualities mapping type IIA supergravity to type IIA so that they appear as a symmetry of the 10-dimensional type IIA supergravity. A single Buscher T-duality maps type IIA to type IIB and thus Buscher T-dualities will always need to be performed in pairs. We will first review the relevant parts of the type IIA double field theory.

\subsection{Double field theory}
Just as for the 11-dimensional SUGRA action, one can rewrite the type IIA 10-dimensional action in a way that makes the T-duality symmetry manifest \cite{Hull:2009zb,Hohm:2010pp,Hull:2009mi,Hohm:2010jy}. When one restricts oneself to act with T-duality in only $n=d-1$ directions,\footnote{Here we use $n=d-1$ to make the connection with $E_{d}$} the T-duality transformations form the group $SO(n,n)$. As mentioned above, a single Buscher T-duality turns type IIA into type IIB and is not a symmetry of the action. These transformations form the parity reversing elements of the full T-duality group, $O(n,n)$, which we ignore in order to make connection with the $E_d$ symmetry group.

The generalised metric unites the string metric $h$ and Kalb-Ramond form $B$ in the $n$ dualisable directions\footnote{We once again ignore fermions and focus only on the bosonic fields, the metric, Kalb-Ramond form and dilaton.} in a T-duality tensor
\begin{equation}
 \gm_S = \left( \begin{array}{cc}
                h - B h^{-1} B & B h^{-1} \\
                - h^{-1} b & h^{-1}
                \end{array} \right).
\end{equation}
We split the indices $\mu, \nu = 1, \ldots 10$ into $i, j = 1, \ldots n$ for the dualisable directions and $A, B = n+1, \ldots 10$ for the undualisable directions. Thus with components written explicitly we have
\begin{equation}
 \left(\gm_S\right)_{MN} = \left( \begin{array}{cc}
                h_{ij} - B_{ik} h^{kl} B_{lj} & B_{ik} h^{kj} \\
                - h^{ik} B_{kj} & h^{ij}
                \end{array} \right).
\end{equation}
Under T-duality the generalised metric obeys a tensor transformation law
\begin{equation}
 \gm_S \rightarrow T^T \gm_S T,
\end{equation}
for $T \in SO(n,n)$, the representation of the T-duality element.

The dilaton, $\phi$, enters through a T-duality scalar
\begin{equation}
 e^{-2d} = \sqrt{|h|} e^{-2\phi},
\end{equation}
i.e.
\begin{equation}
 \phi' = \phi + \frac{1}{4} \ln \frac{|h'|}{|h|}
\end{equation}
under T-duality. Because T-duality exchanges momenta, $P_i$, and winding numbers, $W^i$, of the string
\begin{equation}
 P_i \leftrightarrow W^i,
\end{equation}
the $n$ spacetime coordinates, $x^i$, form a T-duality vector together with coordinates Fourier-dual to the winding numbers, $\tilde{x}_i$,
\begin{equation}
 X^M = \left( \begin{array}{c}
              x^i \\
              \tilde{x}_{i}
              \end{array} \right).
\end{equation}
The T-duality tensors $\gm_S$, $e^{-2d}$ and $X$ can be used to write a manifestly T-duality invariant action for type IIA supergravity just as has been done for the 11-dimensional SUGRA action, equation \eqref{ELEEA}. The assumptions are as before that $g_S$ is factorisable and $B$ has non-vanishing components only along dualisable directions.

\subsection{$SO(n,n)$ T-duality}
The T-duality group $SO(n,n)$ is generated by the elements
\begin{align}
 T_{GL(n)} &= \left( \begin{array}{cc}
                    G & 0 \\
                    0 & G^{-1}
                    \end{array} \right), \\
 T_{b} &= \left( \begin{array}{cc}
                 1 & 0 \\
                 b & 1
                 \end{array} \right), \\
 T_{\beta} &= \left( \begin{array}{cc}
                     1 & \beta \\
                     0 & 1
                     \end{array} \right),
\end{align}
where $G \in GL(n)$, $b_{ij}$ is a two-form and leads to a shift in the Kalb-Ramond form $B_{ij} \rightarrow B_{ij} + b_{ij}$, while $\beta^{ij}$ is a bivector and leads to a shift in the bivector $\gamma^{ij} \rightarrow \gamma^{ij} + \beta^{ij}$.\footnote{Just as in 11-dimensional SUGRA, we can describe the dynamics of type 10-dimensional IIA SUGRA using ($g_S, B$), i.e. a metric and Kalb-Ramond form, or ($\og_S, \gamma$), i.e. a different metric and a bivector \cite{Andriot:2011uh}. The two descriptions are related by a local $O(n)\otimes O(n)$ rotation except in the case of non-geometric backgrounds where the bivector field cannot be gauged away and is related to non-geometric fluxes \cite{Grana:2008yw,Andriot:2011uh,Andriot:2012wx,Andriot:2012an,Aldazabal:2011nj,Dibitetto:2012rk}.}

The $GL(n)$ action can be further decomposed as a scaling, $T_\alpha$, and the action of $SL(n)$, $T_{SL(n)}$,
\begin{align}
 T_{\alpha} &= \left( \begin{array}{cc}
                      \alpha & 0 \\
                      0 & \alpha^{-1}
                      \end{array} \right), \\
 T_{SL(n)} &= \left( \begin{array}{cc}
                     A & 0 \\
                     0 & A^{-1}
                     \end{array} \right),
\end{align}
where $A \in SL(n)$.

The elements $T_{SL(n)}$ generate a $SL(n)$ subgroup in the obvious way, while the quotient group $SO(n,n)/SL(n)$ contains for each of the $\frac{n!}{(n-2)!2!}$ pairs of dualisable directions a $SL(2)$ subgroup generated by
\begin{equation}
 SL(2)_{ij} = \left\{ T_{\alpha}, T_{b_{ij}}, T_{\beta^{ij}} \right\},
\end{equation}
where $i, j$ label the specific pair of directions to be dualised ($i \neq j$). Each $SL(2)_{ij}$ acts on its K\"{a}hler parameter, $E_{(ij)}$,
\begin{equation}
 E_{(ij)} = i \sqrt{|h|} + B_{ij},
\end{equation}
such that
\begin{equation}
 E_{(ij)} \rightarrow \frac{a E_{(ij)} + b}{c E_{(ij)} + d},
\end{equation}
with $ad - bc = 1$. The transformations are related to $T_\alpha, T_b, T_\beta$ analogously as for 11-dimensional supergravity \eqref{EKaehlerA} -- \eqref{EKaehlerO}.

\subsection{$E_{d}$ U-duality and $SO(d-1,d-1)$ T-duality}
11-dimensional supergravity can be reduced along a direction, which we label $\tau$, to obtain the 10-dimensional type IIA supergravity \cite{Witten:1995ex}. The wrapping modes of M2-branes are related to string winding numbers
\begin{equation}
 Z^{\tau i} = W^i,
\end{equation}
and thus the dual coordinates match up as
\begin{equation}
 y_{\tau i} = \tilde{x}_i.
\end{equation}
The 11-dimensional metric gives rise to the string metric, $g_S$, as follows
\begin{equation} \label{EMetricReduction}
 ds_{11}^2 = e^{4\phi/3} (d\tau + \mathcal{A}_\mu dx^{\mu})^2 + e^{-2\phi/3} \left(g_S\right)_{\mu\nu} dx^\mu dx^\nu,
\end{equation}
with indices $\mu, \nu = 1, \ldots 10$ labeling the string theory directions. We further identify $C_{\tau ij} = B_{ij}$, the Kalb-Ramond form of type IIA, and we will also identify $\Omega^{\tau ij} = \beta^{ij}$, the bivector arising in the type IIA double field theory \cite{Grana:2008yw,Andriot:2011uh,Andriot:2012wx,Andriot:2012an,Aldazabal:2011nj,Dibitetto:2012rk}. The $\mathcal{A}_\mu$ are Ramond-Ramond 1-forms while $C_{\mu\nu\gamma}$ are Ramond-Ramond 3-forms.

The $SO(n,n)$ T-duality group is a subgroup of the $E_{d}$ U-duality group. The part of the $SL(d)$ subgroup of $E_{d}$ consisting of interchanges of the dualisable coordinates will map to a $SL(n)$ subgroup of the $SO(n,n)$ T-duality of the corresponding type IIA supergravity. The map is a direct identification of those elements of $SL(d)$ not involving the $\tau$ coordinate to elements in $SL(n)$. The remaining $d-1$ transformations in the $SL(d)$ subgroup of U-duality do not act as T-dualities but rather interchange the string coupling constant $e^{\phi}$ and a string direction. These are expected to give rise to S-duality of the related type IIB supergravity \cite{Schwarz:1995jq,Schwarz:1995dk,Schwarz:1996bh,Vafa:1997pm}.

Of the $\frac{d!}{(d-3)!3!}$ $SL(2)$ subgroups in $E_{d}/SL(d)$ there are $\frac{(d-1)!}{(d-3)!2!}$ involving the coordinate $\tau$. These will map to the $SL(2)$ subgroups of $SO(d-1,d-1)/SL(d-1)$ in a straightforward identification
\begin{align}
 U_{\alpha} & \rightarrow T_{\alpha^{-3}}, \\
 U_{C_{\tau ij}} &\rightarrow T_{B_{ij}}, \\
 U_{\Omega^{\tau ij}} &\rightarrow T_{\beta^{ij}}.
\end{align}
It is trivial to check explicitly that the transformations $U_\alpha$ and $U_{C_{\tau ij}}$ generate the same transformations of $g_S, \phi$ and $B$ as $T_{\alpha^{-3}}$ and $T_{B_{ij}}$.

\subsection{$\Omega$ and $\beta$ shifts}
For $U_{\Omega}$ the calculation is slightly more involved and we will give the details here. We will consider the transformation from the $d = 3$ U-duality group $SL(2) \otimes SL(3)$ which should match a transformation in $SO(2,2)$. The result for $d=4$ is very similar but has the added complication of possible off-diagonal components of the metric and non-trivial transformations of the Ramond-Ramond form. As explained in section \ref{SIntro}, we assume the M-theory metric is factorisable $g_{11} = g_3 \oplus g_8$ and the only non-zero 3-form component is that in the three dualisable directions. This means that the metric ansatz \eqref{EMetricReduction} simplifies to
\begin{equation} \label{EMetricReductionSL5}
 ds^2 = e^{4\phi/3} \left(d\tau + \mathcal{A}_i dx^i \right)^2 + e^{-2\phi/3} \left( h_{ij} dx^i dx^j + G_{AB} dx^A dx^B \right),
\end{equation}
where $i, j = 1, 2$ and $A, B = 3, \ldots 10$ and the string metric $g_S = h \oplus G$ is factorisable into components along the 2 dualisable string theory directions ($i, j = 1, 2$) and the 8 undualisable ones ($A, B = 3, \ldots 10$).

We now dualise along directions $x^\alpha, \alpha = \tau, 1, 2$ and consider $U_\Omega$ with $\Omega^{\tau 12} = A$. We need $|g_3|$, the determinant of the metric in these three directions. We find
\begin{equation}
 \begin{split}
 |g_3| &= g_{\tau\tau} |g_{11}g_{22} - g_{12}^2| = e^{4\phi/3} e^{-4\phi/3} |h_{11} h_{22} - h_{12}^2| \\
 &= |h|.
 \end{split}
\end{equation}
This only works because $g_{\mu\nu} = \frac{1}{\sqrt{g_{\tau\tau}}} h_{\mu\nu}$ which occurs for the reduction ansatz \eqref{EMetricReduction}.

Thus, from \eqref{3OmegaDuality} we get
\begin{align}
 g'_{ij} &= g_{ij} \left( \left(1 + A B_{12} \right)^2 + A^2 |h| \right)^{-2/3}, \\
 g'_{AB} &= g_{AB} \left( \left(1 + A B_{12} \right)^2 + A^2 |h| \right)^{1/3}, \\
 B'_{12} &= \frac{B_{12} \left(1 + A B_{12}\right) + A |h|}{\left(1 + A B_{12} \right)^2 + A^2 |h|}.
\end{align}

Using again the relation between 11-dimensional and 10-dimensional metric, \eqref{EMetricReductionSL5}, we find the new string fields
\begin{align}
 e^{2\phi'/3} &= e^{2\phi/3} \left( \left(1 + A B_{12} \right)^2 + A^2 |h| \right)^{1/3}, \\
 h'_{ij} &= h_{ij} \left( \left(1 + A B_{12} \right)^2 + A^2 |h| \right)^{-1}, \\
 \mathcal{A}'_i &= \mathcal{A}_i, \\
 G'_{AB} &= G_{AB}, \\
 B'_{12} &= \frac{B_{12} \left(1 + A B_{12}\right) + A |h|}{\left(1 + A B_{12} \right)^2 + A^2 |h|}.
\end{align}
This is precisely the $SO(2,2)$ transformation obtained from $T_{\beta}$ with $\beta^{12} = A$.

\subsection{Buscher T-duality}
Because a single Buscher T-duality \cite{Buscher:1987qj,Buscher:1987sk} will turn a type IIA solution into one of type IIB, we will only consider even numbers of Buscher transformations so that we keep IIA supergravity and can make a direct identification with the 11-dimensional supergravity fields. If the duality acted on a two-torus (radii $R_1$ and $R_2$) with vanishing Kalb-Ramond form it would take
\begin{equation}
 \begin{split}
 R_1 & \rightarrow \frac{1}{R_1}, \\
 R_2 & \rightarrow \frac{1}{R_2}.
 \end{split}
\end{equation}

As I will show now the corresponding duality in M-theory is generated by
\begin{equation}
 U_B = \left( \begin{array}{cc}
              0 & \frac{1}{\sqrt{2}} \Omega \\
              \frac{1}{\sqrt{2}} C & 0
              \end{array} \right),
\end{equation}
with $\Omega^{\tau 12} = A, C_{\tau 12} = - \frac{1}{A}$ being the only non-vanishing components of $\Omega$ and $C$. In \eqref{3BuscherDuality} we see the transformation acts as
\begin{equation}
 \begin{split}
  g'_{ij} &= g_{ij} \left( A^2 \left( C_{\tau 12}^2 + |g_3| \right) \right)^{-2/3}, \\
  g'_{AB} &= g_{AB} \left( A^2 \left( C_{\tau 12}^2 + |g_3| \right) \right)^{1/3}, \\
  C'_{\tau 12} &= - \frac{C_{\tau 12}}{A^2 \left( C_{\tau 12}^2 + |g_3| \right)}.
 \end{split}
\end{equation}
For the string variables this implies
\begin{equation}
 \begin{split}
 e^{n \phi'} &= e^{n \phi} \left( A^2 \left( B_{12}^2 + |h| \right) \right)^{1/3}, \\
 h'_{ij} &= h_{ij} \left( A^2 \left( B_{12}^2 + |h| \right) \right)^{-1}, \\
 G'_{AB} &= G_{AB}, \\
 B'_{ij} &= - \frac{B_{12}}{A^2 \left( B_{12}^2 + |h| \right)}.
 \end{split}
\end{equation}
This is generated by the element
\begin{equation}
 \mathcal{T} = \left( \begin{array}{cc}
                    0 & \beta \\
                    b & 0
                    \end{array} \right),
\end{equation}
with $b_{12} = -\beta^{12} = 1$. This is just a Buscher T-duality and an exchange of coordinates. Consider
\begin{align}
 \begin{split}
 T_1 &= \left( \begin{array}{cc}
              m & 1 - m \\
              1 - m & m
              \end{array} \right), \\
 \textrm{where } m &= \left( \begin{array}{cc}
              1 & 0 \\
              0 & 0
              \end{array} \right),
 \end{split} \\
 \begin{split}
 T_{21} &= \left( \begin{array}{cc}
                 0 & n \\
                 n & 0
                 \end{array} \right), \\
 \textrm{where } n &= \left( \begin{array}{cc}
            1 & 0 \\
            0 & 1
            \end{array} \right),
 \end{split} \\
 \begin{split}
 \textrm{and } T_{swap} &= \left( \begin{array}{cc}
                          l & 0 \\
                          0 & l
                          \end{array} \right), \\
 \textrm{where } l & = \left( \begin{array}{cc}
              0 & 1 \\
              -1 & 0
              \end{array} \right).
 \end{split} \\
 \mathcal{T} &= T_{21} \times T_{swap}.
\end{align}
$T_1$ is a Buscher T-duality along a single direction, here $x^1$, while $T_{21}$ is a Buscher T-duality along directions $x^1$ and $x^2$ and is thus an element of $SO(2,2)$. $T_{swap}$ generates an interchange of the $x^1$ and $x^2$ directions. We see that the transformation we have generated is a Buscher duality on two directions and their exchange. This is what $U_B$ corresponds to when $A = 1$. Thus, we see that it generates a Buscher-like T-duality symmetry in the underlying type IIA supergravity.

The 3-torus with vanishing Kalb-Ramond form then transforms under $U_B$ with $\Omega^{123} = 1$ as
\begin{equation} \label{EMBuscher}
 \begin{split}
  R_1 & \rightarrow R_1^{1/3} R_2^{-2/3} R_3^{-2/3}, \\
  R_2 & \rightarrow R_1^{-2/3} R_2^{1/3} R_3^{-2/3} ,\\
  R_3 & \rightarrow R_1^{-2/3} R_2^{-2/3} R_3^{1/3}.
 \end{split}
\end{equation}

It is worth mentioning how the other transformations in the U-duality group are expected to act on the string effective action. There are $\frac{(d-1)!}{(d-4)!3!}$ transformations shifting 3-form components $C_{\mu\nu\gamma}$ that do not involve the $\tau$ direction and the same number of transformations shifting $\Omega_3$ components not involving $\tau$. The former transformations will shift the Ramond-Ramond 3-form of type IIA supergravity. These couple to D2-branes and the $SL(2)$ group involving these transformations, which include the $\Omega$-shift, thus generates dualities of D2-branes. The exact form of these transformations will be discussed in a future paper. The dilatations $U_\alpha$ act similarly but their action is always just a gauge transformation and thus trivial. The remaining part is the $SL(d)$ subgroup which, as mentioned before, generates the $SL(n)$ subgroup of the $SO(n,n)$ T-duality of the type IIA supergravity. The remaining $d-1$ transformations generate S-duality of the related type IIB supergravity \cite{Schwarz:1995jq,Schwarz:1995dk,Schwarz:1996bh,Vafa:1997pm}.

\section{Example of U-duality: M2-brane} \label{SM2Brane}
After this rather abstract exposition of U-dualities, we will now consider some specific examples of the transformations. In the $\left(g_{11}, C_3\right)$ frame, the actions of $U_{SL(d)}$, $U_C$ and $U_\alpha$ are simply gauge transformations. The non-trivial transformation is generated by $U_\Omega$. This transformation always acts in three directions, the ones in which the trivector $\Omega^{ijk}$ is shifted. Under U-duality the generalised coordinates also transform
\begin{equation}
 X^M \rightarrow \left(U^{-1}\right)^M_{\ph{M}N} X^N,
\end{equation}
and specifically for $U_\Omega$ we have
\begin{equation}
 \left( \begin{array}{c}
 x^{i} \\
 \frac{1}{\sqrt{2}} y_{ij} \end{array} \right) \rightarrow
 \left( \begin{array}{c}
 x^i - \frac{1}{2} \Omega^{ijk} y_{jk} \\
 \frac{1}{\sqrt{2}} y_{ij} \end{array} \right).
\end{equation}
The solutions of the low-energy effective action \eqref{ELEEA} that we are used to are those where all fields depend only on spacetime coordinates, $x^a$, and are independent of the dual coordinates, $y_{ij}$. This amounts to a sectioning condition that $\partial_{y}$ acting on all fields, $g_{11}, C_3$ and thus $\gm$ vanishes. To preserve this condition after duality we have to act with $U_\Omega$ only along isometries since
\begin{equation}
 \partial^{ij} \rightarrow \partial'^{ij} = \partial^{ij} + \Omega^{ijk} \partial_{k}.\footnotemark
\end{equation}
\footnotetext{We use the convention that $\partial^{ij} = \frac{\partial}{\partial y_{ij}}$ and $\partial_i = \frac{\partial}{\partial x^i}$.}

The easiest solution with the desired properties would be flat space. It is easy to see that the corresponding transformation will lead to nothing but scalings of the coordinates. The simplest solutions that transform non-trivially then are M2-branes because of their three worldvolume isometries on which we act with the $d=3$ U-duality group $SL(2) \otimes SL(3)$. One could consider $S2$-branes \cite{Gutperle:2002ai} which have three spatial worldvolume directions and a Lorentzian transverse spacetime. These have been studied in the context of Lunin-Maldacena transformations \cite{Lunin:2005jy,CatalOzer:2005mr} in \cite{Kruczenski:2002ap,Deger:2011nb}. However, we will take a different route here and wick-rotate M2-branes to obtain 11-dimensional Euclidean solutions. The Euclidean theory still has the $E_3 \otimes GL(8)$ symmetry as the Lagrangian can be written in a manifestly-invariant form as in equation \eqref{ELEEA}. We use this symmetry to obtain dual 11-dimensonal Euclidean solutions and finally relate these back to Lorentzian solutions by another Wick-rotation. Although the Euclidean solutions are in general complex we will make appropriate coordinate rescalings such that after the final Wick rotation the spacetime is real. These resultant spacetimes are guaranteed to be solutions of the \emph{Lorentzian} 11-dimensional supergravity. While they may not be ``U-dual'' to the original Lorentzian solution, since we obtained them by Wick-rotations and Euclidean U-dualities, one could expect that timelike U-dualities will give similar results. We thus use this procedure to speculate about timelike dualities of Lorentzian M2-branes.

\subsection{Uncharged black M2-brane}
Let's begin with an uncharged black M2-brane \cite{Gueven:1992hh} as this involves no 3-form potential.
\begin{equation}
 \begin{split}
  ds^2 &= -W dt^{2} + dy_1^2 + dy_2^2 + W^{-1} dr^2 + r^2 d\Omega_{(7)}^{2}, \\
  W &= 1 + h/r^{6}, \\
  C_{ty_{1}y_{2}} &= 0,
 \end{split}
\end{equation}
where $r$ is the radius in the six transverse directions, $d\Omega_{(7)}^{2}$\footnote{The symbol $\Omega$ is used here for two different purpose: once in relation to a $S^7$ and once for the trivector. The context will make it clear what is being meant.} corresponds to the metric of a $S^7$ and $\omega_d$ is the volume of a $S^d$
\begin{equation}
 \omega_d = \frac{2\pi^{\frac{d+1}{2}}}{\Gamma(\frac{d+1}{2})}.
\end{equation}

We now Wick rotate to obtain the Euclidean solution.\footnote{As mentioned before we have calculated the action of $E_{d}$ acting on Euclidean spaces in order to avoid the issue of timelike dualities.}
\begin{equation}
  ds^2 = W dt^{2} + dy_1^2 + dy_2^2 + W^{-1} dr^2 + r^2 d\Omega_{(7)}^{2}.
\end{equation}
The coordinates $t, y_1, y_2$ are worldvolume coordinates and form isometries. We will dualise along them with $U_\Omega$, where $\Omega^{ty_1y_2} = A$. Using equations \eqref{3OmegaDuality} we obtain
\begin{equation}
 \begin{split}
  ds'^2 &= \left(1+A^2 W\right)^{-2/3} \left(W dt^2 + dy_1^2 + dy_2^2 \right) + \left(1+A^2 W\right)^{1/3} \left(W^{-1} dr^2 + r^2 d\Omega_{(7)}^{2}\right), \\
  C'_{t y_1 y_2} &= \frac{AW}{1+A^2W}.
 \end{split}
\end{equation}

We first perform a coordinate rescaling so the solution is asymptotically flat in the transverse directions, i.e. we take
\begin{equation}
 \begin{split}
  t & \rightarrow T = t (1+A^2)^{-1/3}, \\
  y_1 & \rightarrow Y_1 = y_1 (1+A^2)^{-1/3}, \\
  y_2 & \rightarrow Y_2 = y_2 (1+A^2)^{-1/3}, \\
  r & \rightarrow R = r (1+A^2)^{1/6}.
 \end{split}
\end{equation}
The metric and 3-form then become
\begin{equation}
 \begin{split}
  ds'^2 &= G^{-2/3} \left(W dT^2 + dY_1^2 + dY_2^2 \right) + G^{1/3} \left(W^{-1} dR^2 + R^2 d\Omega_{(7)}^{2}\right), \\
  C'_{T Y_1 Y_2} &= A - \frac{1}{A} \left(G^{-1} - 1\right),
 \end{split}
\end{equation}
where
\begin{equation}
 \begin{split}
  G &= 1 + \frac{A^2 h}{R^6}, \\
  W &= 1 + \frac{h (1 + A^2)}{R^6}.
 \end{split}
\end{equation}

This is just the supergravity solution describing a Euclidean charged M2-brane. To obtain the Lorentzian solution we Wick rotate back, i.e. $T \rightarrow -i T$. This means we need to take $C_{TY_1Y_2} \rightarrow -i C_{TY_1Y_2}$ and $A \rightarrow i A$. Thus we get
\begin{equation}
 \begin{split}
  ds'^2 &= G^{-2/3} \left(- W dT^2 + dY_1^2 + dY_2^2 \right) + G^{1/3} \left(W^{-1} dR^2 + R^2 d\Omega_{(7)}^{2}\right),\\
  C'_{T Y_1 Y_2} &= A + \frac{1}{A} \left(G^{-1} - 1\right) \label{Harrison},
 \end{split}
\end{equation}
where now
\begin{equation}
 \begin{split}
  G &= 1 - \frac{A^2 h}{R^6},\\
  W &= 1 + \frac{h (1 - A^2)}{R^6}. \label{HarrisonHarmonics}
 \end{split}
\end{equation}

Because the solution is static we can calculate the Komar tension
\begin{equation}
 \begin{split}
  M_2 &= \frac{3}{4\kappa^2} \int_{\partial V} \star K,\footnotemark \\
  \textrm{where } l_a &= g_{ab} \partial_t^b, \\
  K &= dl,
 \end{split}
\end{equation}
and $\partial V$ is the boundary of the spacetime transverse to the membrane. \footnotetext{$\kappa^2 = 8\pi G_N^{(11)}$ where $G_N^{(11)}$ is the 11-dimensional Newton's constant.}

The Page charge \cite{Page:1984qv} of the membrane is given by
\begin{equation}
 \begin{split}
  Q & = \frac{1}{2} \int_{\partial V} \star F + C_3 \wedge F, \\
  \textrm{where } F &= d C_3.
 \end{split}
\end{equation}

The tension of the initial solution is given by
\begin{equation}
 M_2 = -\frac{9 h \omega_7}{2\kappa^2},
\end{equation}
and we find that under the duality the tension and charge density transform as
\begin{equation}
 \begin{split}
  M_2 &\rightarrow M'_2 = \left(1 - \frac{1}{3}A^2 \right) M_2, \\
  Q &\rightarrow Q' = \frac{2}{3} A \kappa^2 M_2 = \frac{2 A \kappa^2 M'_2}{3 - A^2}.
 \end{split}
\end{equation}

Clearly, the uncharged black M2-brane has been dualised into a charged black M2-brane. We have thus taken a vacuum solution of Einstein's equations and transformed it into an electrovac solution. This is an example of a 11-dimensional Harrison transformation \cite{Harrison:1968}. It is important to note that neither the initial nor the transformed solutions are BPS states. Equations \eqref{Harrison} and \eqref{HarrisonHarmonics} show that after Wick-rotating back, the transformed solution will not be real everywhere. In particular, the harmonic function
\begin{equation}
 G = 1 - \frac{A^2h}{R^6}
\end{equation}
becomes negative close to the singularity, while the harmonic function
\begin{equation}
 H = 1 + \frac{h(1-A^2)}{R^6}
\end{equation}
is either positive or negative everywhere, depending on the value of $A$. This is a problem only of the Lorentzian solution obtained by Wick-rotating back and is expected to occur because we have implicitly performed a duality along a timelike direction. We will study this in more detail in a future publication.

Finally, one may want to consider performing a Buscher duality. Using equation \eqref{3BuscherDuality} we find the transformed solution
\begin{equation}
 \begin{split}
  ds^2 & = W^{-2/3} A^{-4/3} \left(W dt^2 + dy_1^2 + dy_2^2\right) + W^{1/3} A^{2/3} \left( W^{-1} dr^2 + r^2 d\Omega_{(7)}^2\right), \\
  C_{ty_1y_2} &= 0,
 \end{split}
\end{equation}
which upon rescaling the coordinates to be asymptotically flat
\begin{equation}
 \begin{split}
 \left( \begin{array}{c} t \\ y_1 \\ y_2 \end{array} \right) &\rightarrow \left( \begin{array}{c} T \\ Y_1 \\ Y_2 \end{array} \right) = A^{-1/3} \left( \begin{array}{c} t \\ y_1 \\ y_2 \end{array} \right), \\
 r & \rightarrow R = r A^{1/6},
 \end{split}
\end{equation}
and Wick-rotating as before becomes
\begin{equation}
 \begin{split}
  ds^2 & = W^{-2/3} \left(-W dT^2 + dY_1^2 + dY_2^2\right) + W^{1/3} \left( W^{-1} dR^2 + R^2 d\Omega_{(7)}^2\right), \\
  C_{TY_1Y_2} &= 0,
 \end{split}
\end{equation}
where $W = 1 - \frac{h A^2}{R^6}$. This is again an uncharged black M2-brane with tension given by
\begin{equation}
 M'_2 = \frac{3A^2}{2\kappa^2} h \omega_7 = - \frac{A^2}{3}M_2.
\end{equation}
Once again the Lorentzian continuation has a harmonic function that is negative close to the singularity because of the implicit timelike duality. These problems are likely to be related to the difficulties associated with timelike dualities as in \cite{Hull:1998vg,Hull:1998ym}. The Euclidean dual solution does not, however, exhibit any such problems.

\subsection{Extreme M2-Branes}
Another example of interest is dualising an extreme M2-brane \cite{Duff:1990xz}. This a 1/2-BPS state and thus its tension and charge are protected from quantum effects. The Euclidean supergravity solution has metric and 3-form given by\footnote{This can be obtained from the usual solution by Wick-rotating. This gives rise to the imaginary charge as $C_{ty_1y_2} \rightarrow i C_{ty_1y_2}$}
\begin{equation}
 \begin{split}
  ds^2 &= H^{-2/3} \left(dt^2 + dy_1^2 + dy_2^2\right) + H^{1/3} \left(dr^2 + r^2 d\Omega_{(7)}^{2}\right), \\
  C_{ty_1y_2} & = \pm i H^{-1}, \\
  H &= 1 + \frac{h}{r^6}.
 \end{split}
\end{equation}

We again act with $\Omega^{ty_1y_2} = A$ to obtain
\begin{equation}
 \begin{split}
  ds'^2 &= (fH)^{-2/3} \left(dt^2 + dy_1^2 + dy_2^2\right) + (fH)^{1/3} \left(dr^2 + r^2 d\Omega_{(7)}^{2}\right), \\
  C'_{ty_1y_2} &= \pm i \left(fH\right)^{-1}, \\
  f &= 1 \pm 2i A H^{-1}.
 \end{split}
\end{equation}
Rescaling the coordinates once again to make the solution asymptotically flat, $(t, y_1, y_2) \rightarrow (1\pm2iA)^{-1/3} (t, y_1, y_2)$ and $r \rightarrow r (1\pm2iA)^{1/6}$ we have
\begin{equation}
 \begin{split}
  ds'^2 &= G^{-2/3} \left(dT^2 + dY_1^2 + dY_2^2\right) + G^{1/3} \left(dR^2 + R^2 d\Omega_{(7)}^{2}\right), \\
  C'_{TY_1Y_2} &= \pm i G^{-1}, \\
  G &= 1 + \frac{h}{R^6}
 \end{split}
\end{equation}
and so we see that we obtain the same extreme M2-brane solution. Therefore under the continuous duality, $U_{\Omega}$, the extreme M2-brane is self-dual.

\subsubsection{Extreme M2-brane with a C-shift}
Although the extreme M2-brane is self-dual in the parametrisation used above, we can consider performing a gauge transformation of the 3-form potential $C_3$ before dualising. So we take
\begin{equation}
 \begin{split}
  ds^2 &= H^{-2/3} \left(dt^2 + dy_1^2 + dy_2^2\right) + H^{1/3} \left(dr^2 + r^2 d\Omega_{(7)}^{2}\right), \\
  C_{ty_1y_2} & = n \pm i H^{-1}, \\
  H &= 1 + \frac{h}{r^6}.
 \end{split}
\end{equation}
We consider $n = const.$ here so that the gauge transformation is actually generated by a $C$-shift, $U_{C} \in E_{3}$, with $C_{ty_1y_2}=n$.

Performing a duality transformation $U_{\Omega}$, where $\Omega^{ty_1y_2}=A$, we now obtain
\begin{equation}
 \begin{split}
  ds'^2 &= (fH)^{-2/3} \left(dt^2 + dy_1^2 + dy_2^2\right) + (fH)^{1/3} \left(dr^2 + r^2 d\Omega_{(7)}^{2}\right) \\
  C'_{ty_1y_2} &= \frac{n\left(1+An\right) \pm i \left( 1 + 2 An\right) + n \left(1+An\right) \frac{h}{r^6}}{fH} \\
  f &= \left(1+An\right) \left(1+An \pm 2i A H^{-1}\right)
 \end{split}
\end{equation}
so that we rescale the coordinates
\begin{equation}
 \begin{split}
  \left( \begin{array}{c}
  t \\ y_1 \\ y_2 \end{array} \right) &\rightarrow \left[\left(1+An\right)\left(1+An\pm2iA\right)\right]^{-1/3}
  \left( \begin{array}{c}
  t \\ y_1 \\ y_2 \end{array} \right), \\
  r & \rightarrow r \left[ \left(1+An\right)\left(1+An\pm2iA\right)\right]^{1/6},
  \end{split}
\end{equation} in order to obtain an asymptotically flat metric.
\begin{equation}
 \begin{split}
  ds'^2 &= G^{-2/3} \left(dT^2 + dY_1^2 + dY_2^2\right) + G^{1/3} \left(dR^2 + R^2 d\Omega_{(7)}^{2}\right), \\
  C'_{TY_1Y_2} &= \pm i G^{-1} + n \left(1+An\right)\left(1\pm2iAn\right), \\
  G &= 1 + \left(1+An\right)^2 \frac{h}{R^6}.
 \end{split}
\end{equation}

Wick-rotating back\footnote{$n \rightarrow -in$ here as well as it is part of the $C_{ty_1y_2}$ component}
\begin{equation}
 \begin{split}
  ds'^2 &= G^{-2/3} \left(-dT^2 + dY_1^2 + dY_2^2\right) + G^{1/3} \left(dR^2 + R^2 d\Omega_{(7)}^{2}\right), \\
  C'_{TY_1Y_2} &= \pm G^{-1} - n \left(1+An\right)\left(1\pm2An\right), \\
  G &= 1 + \left(1+An\right)^2 \frac{h}{R^6}.
 \end{split}
\end{equation}
This solution describes another extreme membrane with tension and charge density given by
\begin{equation}
 \begin{split}
  M'_2 &= M_2 \left(1 + An \right)^2, \\
  Q' &= Q \left( 1 + An \right)^2,
 \end{split}
\end{equation}
so the duality transformation turns the extreme M2-brane into one with different tension and charge density.

\subsubsection{Buscher transformation of an extreme M2-brane}
If, on the other hand, we perform a Buscher-duality we find a singular transformation in equation \eqref{3BuscherDuality} since
\begin{equation}
 |g| = H^{-2} = - C_{ty_1y_2}^2
\end{equation}
in the Wick-rotated solution.
Thus, from equation \eqref{3BuscherDuality} it would seem that one cannot perform a Buscher duality. However, if we remember that a Buscher transformation, $U_B$, can be generated by the three successive transformations (c.f. equation \eqref{EBuscherGenerated})
\begin{equation}
 U_B = U_C U_\Omega U_C,
\end{equation}
where $C_{ty_1y_2} = -\frac{1}{A}$ and $\Omega^{ty_1y_2} = A$, then we naively would expect to be able to use the result from the previous result with $n = -\frac{1}{A}$. In that case we seem to be getting flat space since
\begin{equation}
 1 + An = 1 - 1 = 0,
\end{equation}
and so
\begin{equation}
 G = 1 + \frac{ 0 \times h }{R^6}.
\end{equation}
However, it would be very surprising if it were true that the extreme M2-brane can be Buscher dualised to flat space: since the transformation is invertible, it would mean that flat space in the appropriate frame can be Buscher dualised to an extreme M2-brane.

One can see what goes wrong in the argument if we add a small parameter, $\epsilon$, to the 3-form which regularises the singularity in the Buscher transformation
\begin{equation}
 C_{ty_1y_2} = i H^{-1} + \epsilon.
\end{equation}
Then one finds that to $O(\epsilon)$ the resultant Euclidean dual solution is
\begin{equation}
 \begin{split}
 ds'^2 &= \epsilon^{-2/3} A^{-4/3} \left(H \epsilon \pm 2i \right)^{-2/3} \left( dt^2 + dy_1^2 + dy_2^2 \right) + \epsilon^{1/3} A^{2/3} \left(H \epsilon \pm 2i \right)^{1/3} \left( dr^2 + r^2 d\Omega_{(7)}^2 \right), \\
 C'_{ty_1y_2} &= \mp i \frac{1}{A^2 \epsilon} \frac{1}{H \epsilon \pm 2i},
 \end{split}
\end{equation}
and by rescaling coordinates so the metric becomes manifestly asymptotically flat, i.e. $(t, y_1, y_2) \rightarrow (T, Y_1, Y_2) = \left( \pm 2i \epsilon A^2 \right)^{-1/3}$ and $r \rightarrow R = r (\pm 2i \epsilon A^2)^{1/6}$ to $O(\epsilon)$, we obtain
\begin{equation}
 \begin{split}
 ds'^2 &= G^{-2/3} \left(dT^2 + dY_1^2 + dY_2^2\right) + G^{1/3}\left(dR^2 + R^2 d\Omega_{(7)}^2\right), \\
 C' &= \mp i G^{-1}, \\
 G &= 1 \pm \frac{\epsilon}{2i} + \epsilon^2 A^2 \frac{h}{R^6}.
 \end{split}
\end{equation}
To leading order in $\epsilon$ we can take $G$ to be
\begin{equation}
 G = 1 + \epsilon^2 A \frac{h}{R^6}
\end{equation}
but clearly the $\epsilon^2$ term can only be ignored where $R \neq 0$ that is away from the coordinate singularity at $R=0$. In the naive result obtained by using the results from the previous section, this is ignored which is why the dual solution seems flat. The final interpretation of this issue is not yet clear but it once again seems to be related to the difficulty of performing dualities in timelike directions.

\subsubsection{Quantum effects}

Because the M2-brane is a 1/2-BPS state it allows us to have a glimpse at the quantum theory. We expect that quantum effects break the U-duality group down into its discrete part $E_{d}(Z)$, for example $SL(2,Z) \times SL(3,Z)$ in the 3-d case, or $SL(5,Z)$ in the 4-d case \cite{Hull:1994ys}. This would mean that the parameters $A$ and $n$ are integer valued. The consequence for the transformation would be that the new tensions and charges are just integer multiples of the old ones. Since the extreme M2-brane is a fundamental object of the quantum theory, any extreme solution must have a tension proportional to the fundamental quantum of membrane tension, $M_2 = N M_{2,\textrm{fundamental}}$,\footnote{$M_{2,fundamental} = \frac{l^{6}}{\kappa^2}$ and $l$ is the Planck length.} corresponding to $N$ fundamental membranes stacked on top of each other. A U-duality transformation then gives a new extreme M2-brane solution of tension $M'_2 = (1+An)M_2$, that is a solution corresponding to $(1+An)N$ M2-branes stacked upon each other. We see that the discrete U-duality groups preserve mass (tension) and charge (density) quantisation.

\section{Conclusion}
We have seen how the 11-dimensional action can be written in a way that makes its U-duality symmetry manifest. For $d <5$ the U-duality group's non-trivial transformation is generated by the trivector shifts, $U_\Omega$, with the other transformations being diffeomorphisms and $C_3$ gauge transformations. We expect a similar structure in higher dimensional duality groups which we will study in future papers.

The M2-brane is a good arena to study U-dualities as its three worldvolume directions are isometries. However, no compactification is imposed so that the notion of U-duality used here goes beyond that usually seen in the literature. Here, it simply means the action of $E_{d}$ on the supergravity solution. We found that dualities transform uncharged black M2-branes into charged black ones while the the extreme M2-brane is self-dual. However, by acting with gauge transformations before dualising the 1/2-BPS solution one can obtain a non-trivial transformation. This generates new extreme M2-branes with tensions and charges proportional to the original ones. The proportionality factor becomes an integer when using the discrete U-duality group $E_d(Z)$. This means that charge and mass quantisation is preserved under discrete U-duality.

Finally, it is worth mentioning that the formulae presented here are for Euclidean directions which is why we Wick-rotated the solutions. In the Lorentzian case, the U-duality group $E_d$ remains the same but the local symmetry group $H_d$ is no longer the maximal compact subgroup as given in table \ref{TDualityGroups}, but rather a non-compact subgroup of $E_d$ \cite{Hull:1998br}. This is similar to the case of pure gravity which can be viewed as a non-linear realisation of $GL(d)$ \cite{Borisov:1974bn} with local symmetry group $SO(d)$ for $d$ Euclidean directions. In the Lorentzian case, the rigid diffeomorphism group is still $GL(d)$ but the local symmetry group becomes the Lorentz group $SO(d-1,1)$. Thus it should not come as a surprise that the U-duality group remains the same and only the local symmetries change as has been argued from a group-theoretical perspective in \cite{Hull:1998br}. We can then follow the same procedure as in this paper for the Lorentzian case. One finds similar results but with the equations of section \ref{SOmega} differing. In fact they differ exactly by what we found when Wick-rotating the Euclidean duality transformations in section \ref{SM2Brane}, that is $C_{ty_1y_2}^2 \rightarrow - C_{ty_1y_2}^2$. We also find an obstruction to these dualities which may avoid the difficulties arising for time-like dualities found in \cite{Hull:1998vg,Hull:1998ym} and which have been partially seen in the preceding section. We will return to this issue in a future publication.

\acknowledgments
I would like to thank my supervisor, Malcolm Perry, for many helpful discussions, the STFC for supporting me through a Postgraduate Studentship grant and Peterhouse, Cambridge, for their support through the Peterhouse Research Studentship.

\bibliographystyle{JHEP}
\bibliography{Bibliography}

\providecommand{\href}[2]{#2}\begingroup\raggedright\begin{thebibliography}{10%
0}

\bibitem{Cremmer:1977tt}
E.~Cremmer, J.~Scherk, and S.~Ferrara, {\it {SU(4) Invariant Supergravity
  Theory}},  {\em Phys.\ Lett.} {\bf B 74} (1978) 61.

\bibitem{Cremmer:1978ds}
E.~Cremmer and B.~Julia, {\it {The N=8 Supergravity Theory. 1. The
  Lagrangian}},  {\em Phys.\ Lett.} {\bf B 80} (1978) 48.

\bibitem{Cremmer:1979up}
E.~Cremmer and B.~Julia, {\it {The SO(8) Supergravity}},  {\em Nucl.\ Phys.}
  {\bf B 159} (1979) 141.

\bibitem{Buscher:1987sk}
T.~H. Buscher, {\it {A Symmetry of the String Background Field Equations}},
  {\em Phys.\ Lett.} {\bf B 194} (1987) 59.

\bibitem{Buscher:1987qj}
T.~H. Buscher, {\it {Path Integral Derivation of Quantum Duality in Nonlinear
  Sigma Models}},  {\em Phys.\ Lett.} {\bf B 201} (1988) 466.

\bibitem{Nicolai:1986jk}
H.~Nicolai, {\it {D = 11 supergravity with local SO(16) invariance}},  {\em
  Phys.Lett.} {\bf B187} (1987) 316.

\bibitem{Duff:1985bv}
M.~J. Duff, {\it {E8 $\times$ SO(16) Symmetry of d = 11 Supergravity}}, .

\bibitem{deWit:1986mz}
B.~de~Wit and H.~Nicolai, {\it {d = 11 supergravity with local SU(8)
  invariance}},  {\em Nucl.\ Phys.} {\bf B 274} (1986) 363.

\bibitem{deWit:2000wu}
B.~de~Wit and H.~Nicolai, {\it {Hidden symmetries, central charges and all
  that}},  {\em Class.Quant.Grav.} {\bf 18} (2001) 3095--3112,
  [\href{http://xxx.lanl.gov/abs/hep-th/0011239}{{\tt hep-th/0011239}}].

\bibitem{West:2010rv}
P.~C. West, {\it {Generalised space-time and duality}},  {\em Phys.Lett.} {\bf
  B693} (2010) 373--379, [\href{http://xxx.lanl.gov/abs/1006.0893}{{\tt
  arXiv:1006.0893}}].

\bibitem{West:2001as}
P.~C. West, {\it {E(11) and M theory}},  {\em Class.\ Quant.\ Grav.} {\bf 18}
  (2001) 4443--4460, [\href{http://xxx.lanl.gov/abs/hep-th/0104081}{{\tt
  hep-th/0104081}}].

\bibitem{West:2003fc}
P.~C. West, {\it {E(11), SL(32) and central charges}},  {\em Phys.Lett.} {\bf
  B575} (2003) 333--342, [\href{http://xxx.lanl.gov/abs/hep-th/0307098}{{\tt
  hep-th/0307098}}].

\bibitem{West:2004kb}
P.~C. West, {\it {E(11) origin of brane charges and U-duality multiplets}},
  {\em JHEP} {\bf 0408} (2004) 052,
  [\href{http://xxx.lanl.gov/abs/hep-th/0406150}{{\tt hep-th/0406150}}].

\bibitem{Kleinschmidt:2003jf}
A.~Kleinschmidt and P.~C. West, {\it {Representations of G+++ and the role of
  space-time}},  {\em JHEP} {\bf 0402} (2004) 033,
  [\href{http://xxx.lanl.gov/abs/hep-th/0312247}{{\tt hep-th/0312247}}].

\bibitem{West:2004iz}
P.~C. West, {\it {Brane dynamics, central charges and E(11)}},  {\em JHEP} {\bf
  0503} (2005) 077, [\href{http://xxx.lanl.gov/abs/hep-th/0412336}{{\tt
  hep-th/0412336}}].

\bibitem{West:2011mm}
P.~C. West, {\it {Generalised geometry, eleven dimensions and E11}},  {\em
  JHEP} {\bf 1202} (2012) 018, [\href{http://xxx.lanl.gov/abs/1111.1642}{{\tt
  arXiv:1111.1642}}].

\bibitem{Riccioni:2007ni}
F.~Riccioni and P.~C. West, {\it {E(11)-extended spacetime and gauged
  supergravities}},  {\em JHEP} {\bf 0802} (2008) 039,
  [\href{http://xxx.lanl.gov/abs/0712.1795}{{\tt arXiv:0712.1795}}].

\bibitem{Nicolai:2010zz}
H.~Nicolai and A.~Kleinschmidt, {\it {E10: A fundamental symmetry in
  physics?}},  {\em Phys.Unserer Zeit} {\bf 3N41} (2010) 134--140.

\bibitem{Damour:2002cu}
T.~Damour, M.~Henneaux, and H.~Nicolai, {\it {E(10) and a 'small tension
  expansion' of M theory}},  {\em Phys.Rev.Lett.} {\bf 89} (2002) 221601,
  [\href{http://xxx.lanl.gov/abs/hep-th/0207267}{{\tt hep-th/0207267}}].

\bibitem{Hillmann:2009ci}
C.~Hillmann, {\it {Generalized E(7(7)) coset dynamics and D=11 supergravity}},
  {\em JHEP} {\bf 0903} (2009) 135,
  [\href{http://xxx.lanl.gov/abs/0901.1581}{{\tt arXiv:0901.1581}}].

\bibitem{Gualtieri:2003dx}
M.~Gualtieri, {\it {Generalized complex geometry}},
  \href{http://xxx.lanl.gov/abs/math/0401221}{{\tt math/0401221}}.

\bibitem{Hitchin:2004ut}
N.~Hitchin, {\it {Generalized Calabi-Yau manifolds}},  {\em Quart.\ J.\ Math.\
  Oxford Ser.} {\bf 54} (2003) 281--308,
  [\href{http://xxx.lanl.gov/abs/math/0209099}{{\tt math/0209099}}].

\bibitem{Hitchin:2005in}
N.~Hitchin, {\it {Brackets, forms and invariant functionals}},
  \href{http://xxx.lanl.gov/abs/math/0508618}{{\tt math/0508618}}.

\bibitem{Hitchin:2005cv}
N.~Hitchin, {\it {Instantons, Poisson structures and generalized Kahler
  geometry}},  {\em Commun.\ Math.\ Phys.} {\bf 265} (2006) 131--164,
  [\href{http://xxx.lanl.gov/abs/math/0503432}{{\tt math/0503432}}].

\bibitem{Hull:2007zu}
C.~M. Hull, {\it {Generalised Geometry for M-Theory}},  {\em JHEP} {\bf 0707}
  (2007) 079, [\href{http://xxx.lanl.gov/abs/hep-th/0701203}{{\tt
  hep-th/0701203}}].

\bibitem{Berman:2010is}
D.~S. Berman and M.~J. Perry, {\it {Generalized Geometry and M theory}},  {\em
  JHEP} {\bf 1106} (2011) 074, [\href{http://xxx.lanl.gov/abs/1008.1763}{{\tt
  arXiv:1008.1763}}].

\bibitem{Berman:2011pe}
D.~S. Berman, H.~Godazgar, and M.~J. Perry, {\it {SO(5,5) duality in M-theory
  and generalized geometry}},  {\em Phys.\ Lett.} {\bf B 700} (2011) 65--67,
  [\href{http://xxx.lanl.gov/abs/1103.5733}{{\tt arXiv:1103.5733}}].

\bibitem{Berman:2011jh}
D.~S. Berman, H.~Godazgar, M.~J. Perry, and P.~C. West, {\it {Duality Invariant
  Actions and Generalised Geometry}},  {\em JHEP} {\bf 1202} (2012) 108,
  [\href{http://xxx.lanl.gov/abs/1111.0459}{{\tt arXiv:1111.0459}}].

\bibitem{Berman:2011cg}
D.~S. Berman, H.~Godazgar, M.~Godazgar, and M.~J. Perry, {\it {The Local
  symmetries of M-theory and their formulation in generalised geometry}},  {\em
  JHEP} {\bf 1201} (2012) 012, [\href{http://xxx.lanl.gov/abs/1110.3930}{{\tt
  arXiv:1110.3930}}].

\bibitem{Hull:2009zb}
C.~M. Hull and B.~Zwiebach, {\it {The Gauge algebra of double field theory and
  Courant brackets}},  {\em JHEP} {\bf 0909} (2009) 090,
  [\href{http://xxx.lanl.gov/abs/0908.1792}{{\tt arXiv:0908.1792}}].

\bibitem{Hull:2009mi}
C.~M. Hull and B.~Zwiebach, {\it {Double Field Theory}},  {\em JHEP} {\bf 0909}
  (2009) 099, [\href{http://xxx.lanl.gov/abs/0904.4664}{{\tt
  arXiv:0904.4664}}].

\bibitem{Hohm:2010pp}
O.~Hohm, C.~M. Hull, and B.~Zwiebach, {\it {Generalized metric formulation of
  double field theory}},  {\em JHEP} {\bf 1008} (2010) 008,
  [\href{http://xxx.lanl.gov/abs/1006.4823}{{\tt arXiv:1006.4823}}].

\bibitem{Hohm:2010jy}
O.~Hohm, C.~M. Hull, and B.~Zwiebach, {\it {Background independent action for
  double field theory}},  {\em JHEP} {\bf 1007} (2010) 016,
  [\href{http://xxx.lanl.gov/abs/1003.5027}{{\tt arXiv:1003.5027}}].

\bibitem{Hull:2006tp}
C.~M. Hull and R.~A. Reid-Edwards, {\it {Flux compactifications of M-theory on
  twisted Tori}},  {\em JHEP} {\bf 0610} (2006) 086,
  [\href{http://xxx.lanl.gov/abs/hep-th/0603094}{{\tt hep-th/0603094}}].

\bibitem{Pacheco:2008ps}
P.~P. Pacheco and D.~Waldram, {\it {M-theory, exceptional generalised geometry
  and superpotentials}},  {\em JHEP} {\bf 0809} (2008) 123,
  [\href{http://xxx.lanl.gov/abs/0804.1362}{{\tt arXiv:0804.1362}}].

\bibitem{Aldazabal:2010ef}
G.~Aldazabal, E.~Andres, P.~G. Camara, and M.~Gra\~{n}a, {\it {U-dual fluxes
  and Generalized Geometry}},  {\em JHEP} {\bf 1011} (2010) 083,
  [\href{http://xxx.lanl.gov/abs/1007.5509}{{\tt arXiv:1007.5509}}].

\bibitem{Coimbra:2011ky}
A.~Coimbra, C.~Strickland-Constable, and D.~Waldram, {\it {$E_{d(d)} \times
  R^+$ Generalised Geometry, Connections and M Theory}},
  \href{http://xxx.lanl.gov/abs/1112.3989}{{\tt arXiv:1112.3989}}.

\bibitem{Hull:2004in}
C.~M. Hull, {\it {A Geometry for non-geometric string backgrounds}},  {\em
  JHEP} {\bf 0510} (2005) 065,
  [\href{http://xxx.lanl.gov/abs/hep-th/0406102}{{\tt hep-th/0406102}}].

\bibitem{Hull:2005hk}
C.~M. Hull and R.~A. Reid-Edwards, {\it {Flux compactifications of string
  theory on twisted tori}},  {\em Fortsch.\ Phys.} {\bf 57} (2009) 862--894,
  [\href{http://xxx.lanl.gov/abs/hep-th/0503114}{{\tt hep-th/0503114}}].

\bibitem{Dabholkar:2005ve}
A.~Dabholkar and C.~M. Hull, {\it {Generalised T-duality and non-geometric
  backgrounds}},  {\em JHEP} {\bf 0605} (2006) 009,
  [\href{http://xxx.lanl.gov/abs/hep-th/0512005}{{\tt hep-th/0512005}}].

\bibitem{Hull:2006va}
C.~M. Hull, {\it {Doubled Geometry and T-Folds}},  {\em JHEP} {\bf 0707} (2007)
  080, [\href{http://xxx.lanl.gov/abs/hep-th/0605149}{{\tt hep-th/0605149}}].

\bibitem{Grana:2008yw}
M.~Gra\~{n}a, R.~Minasian, M.~Petrini, and D.~Waldram, {\it {T-duality,
  Generalized Geometry and Non-Geometric Backgrounds}},  {\em JHEP} {\bf 0904}
  (2009) 075, [\href{http://xxx.lanl.gov/abs/0807.4527}{{\tt
  arXiv:0807.4527}}].

\bibitem{Hull:2009sg}
C.~M. Hull and R.~A. Reid-Edwards, {\it {Non-geometric backgrounds, doubled
  geometry and generalised T-duality}},  {\em JHEP} {\bf 0909} (2009) 014,
  [\href{http://xxx.lanl.gov/abs/0902.4032}{{\tt arXiv:0902.4032}}].

\bibitem{Aldazabal:2011nj}
G.~Aldazabal, W.~Baron, D.~Marques, and C.~Nunez, {\it {The effective action of
  Double Field Theory}},  {\em JHEP} {\bf 1111} (2011) 052,
  [\href{http://xxx.lanl.gov/abs/1109.0290}{{\tt arXiv:1109.0290}}].

\bibitem{Grana:2012rr}
M.~Gra\~{n}a and D.~Marques, {\it {Gauged Double Field Theory}},  {\em JHEP}
  {\bf 1204} (2012) 020, [\href{http://xxx.lanl.gov/abs/1201.2924}{{\tt
  arXiv:1201.2924}}].

\bibitem{Dibitetto:2012rk}
G.~Dibitetto, J.~J. Fernandez-Melgarejo, D.~Marques, and D.~Roest, {\it
  {Duality orbits of non-geometric fluxes}},
  \href{http://xxx.lanl.gov/abs/1203.6562}{{\tt arXiv:1203.6562}}.

\bibitem{Andriot:2012wx}
D.~Andriot, O.~Hohm, M.~Larfors, D.~Lust, and P.~Patalong, {\it {A geometric
  action for non-geometric fluxes}},
  \href{http://xxx.lanl.gov/abs/1202.3060}{{\tt arXiv:1202.3060}}.

\bibitem{Andriot:2012an}
D.~Andriot, O.~Hohm, M.~Larfors, D.~Lust, and P.~Patalong, {\it {Non-Geometric
  Fluxes in Supergravity and Double Field Theory}},
  \href{http://xxx.lanl.gov/abs/1204.1979}{{\tt arXiv:1204.1979}}.

\bibitem{Jeon:2010rw}
I.~Jeon, K.~Lee, and J.-H. Park, {\it {Differential geometry with a projection:
  Application to double field theory}},  {\em JHEP} {\bf 1104} (2011) 014,
  [\href{http://xxx.lanl.gov/abs/1011.1324}{{\tt arXiv:1011.1324}}].

\bibitem{Jeon:2011kp}
I.~Jeon, K.~Lee, and J.-H. Park, {\it {Double field formulation of Yang-Mills
  theory}},  {\em Phys.Lett.} {\bf B701} (2011) 260--264,
  [\href{http://xxx.lanl.gov/abs/1102.0419}{{\tt arXiv:1102.0419}}].

\bibitem{Jeon:2011cn}
I.~Jeon, K.~Lee, and J.-H. Park, {\it {Stringy differential geometry, beyond
  Riemann}},  {\em Phys.Rev.} {\bf D84} (2011) 044022,
  [\href{http://xxx.lanl.gov/abs/1105.6294}{{\tt arXiv:1105.6294}}].

\bibitem{Jeon:2011vx}
I.~Jeon, K.~Lee, and J.-H. Park, {\it {Incorporation of fermions into double
  field theory}},  {\em JHEP} {\bf 1111} (2011) 025,
  [\href{http://xxx.lanl.gov/abs/1109.2035}{{\tt arXiv:1109.2035}}].

\bibitem{Jeon:2011sq}
I.~Jeon, K.~Lee, and J.-H. Park, {\it {Supersymmetric Double Field Theory:
  Stringy Reformulation of Supergravity}},  {\em Phys.Rev.} {\bf D85} (2012)
  081501, [\href{http://xxx.lanl.gov/abs/1112.0069}{{\tt arXiv:1112.0069}}].

\bibitem{Hohm:2011ex}
O.~Hohm and S.~K. Kwak, {\it {Double Field Theory Formulation of Heterotic
  Strings}},  {\em JHEP} {\bf 1106} (2011) 096,
  [\href{http://xxx.lanl.gov/abs/1103.2136}{{\tt arXiv:1103.2136}}].

\bibitem{Hohm:2011zr}
O.~Hohm, S.~K. Kwak, and B.~Zwiebach, {\it {Unification of Type II Strings and
  T-duality}},  {\em Phys.\ Rev.\ Lett.} {\bf 107} (2011) 171603,
  [\href{http://xxx.lanl.gov/abs/1106.5452}{{\tt arXiv:1106.5452}}].

\bibitem{Hohm:2011dv}
O.~Hohm, S.~K. Kwak, and B.~Zwiebach, {\it {Double Field Theory of Type II
  Strings}},  {\em JHEP} {\bf 1109} (2011) 013,
  [\href{http://xxx.lanl.gov/abs/1107.0008}{{\tt arXiv:1107.0008}}].

\bibitem{Hohm:2011cp}
O.~Hohm and S.~K. Kwak, {\it {Massive Type II in Double Field Theory}},  {\em
  JHEP} {\bf 1111} (2011) 086, [\href{http://xxx.lanl.gov/abs/1108.4937}{{\tt
  arXiv:1108.4937}}].

\bibitem{Hohm:2011nu}
O.~Hohm and S.~K. Kwak, {\it {N=1 Supersymmetric Double Field Theory}},  {\em
  JHEP} {\bf 1203} (2012) 080, [\href{http://xxx.lanl.gov/abs/1111.7293}{{\tt
  arXiv:1111.7293}}].

\bibitem{Hohm:2011si}
O.~Hohm and B.~Zwiebach, {\it {On the Riemann Tensor in Double Field Theory}},
  {\em JHEP} {\bf 1205} (2012) 126,
  [\href{http://xxx.lanl.gov/abs/1112.5296}{{\tt arXiv:1112.5296}}].

\bibitem{ReidEdwards:2010vp}
R.~A. Reid-Edwards, {\it {Bi-Algebras, Generalised Geometry and T-Duality}},
  \href{http://xxx.lanl.gov/abs/1001.2479}{{\tt arXiv:1001.2479}}.

\bibitem{Coimbra:2011nw}
A.~Coimbra, C.~Strickland-Constable, and D.~Waldram, {\it {Supergravity as
  Generalised Geometry I: Type II Theories}},  {\em JHEP} {\bf 1111} (2011)
  091, [\href{http://xxx.lanl.gov/abs/1107.1733}{{\tt arXiv:1107.1733}}].

\bibitem{Coimbra:2012yy}
A.~Coimbra, C.~Strickland-Constable, and D.~Waldram, {\it {Generalised Geometry
  and type II Supergravity}},  \href{http://xxx.lanl.gov/abs/1202.3170}{{\tt
  arXiv:1202.3170}}.

\bibitem{Albertsson:2008gq}
C.~Albertsson, T.~Kimura, and R.~A. Reid-Edwards, {\it {D-branes and doubled
  geometry}},  {\em JHEP} {\bf 0904} (2009) 113,
  [\href{http://xxx.lanl.gov/abs/0806.1783}{{\tt arXiv:0806.1783}}].

\bibitem{Albertsson:2011ux}
C.~Albertsson, S.-H. Dai, P.-W. Kao, and F.-L. Lin, {\it {Double Field Theory
  for Double D-branes}},  {\em JHEP} {\bf 1109} (2011) 025,
  [\href{http://xxx.lanl.gov/abs/1107.0876}{{\tt arXiv:1107.0876}}].

\bibitem{Copland:2011wx}
N.~B. Copland, {\it {A Double Sigma Model for Double Field Theory}},  {\em
  JHEP} {\bf 1204} (2012) 044, [\href{http://xxx.lanl.gov/abs/1111.1828}{{\tt
  arXiv:1111.1828}}].

\bibitem{Copland:2011yh}
N.~B. Copland, {\it {Connecting T-duality invariant theories}},  {\em
  Nucl.Phys.} {\bf B854} (2012) 575--591,
  [\href{http://xxx.lanl.gov/abs/1106.1888}{{\tt arXiv:1106.1888}}].

\bibitem{Kan:2011vg}
N.~Kan, K.~Kobayashi, and K.~Shiraishi, {\it {Equations of Motion in Double
  Field Theory: From particles to scale factors}},  {\em Phys.Rev.} {\bf D84}
  (2011) 124049, [\href{http://xxx.lanl.gov/abs/1108.5795}{{\tt
  arXiv:1108.5795}}].

\bibitem{Roiban:2012gi}
R.~Roiban and A.~A. Tseytlin, {\it {On duality symmetry in perturbative quantum
  theory}},  \href{http://xxx.lanl.gov/abs/1205.0176}{{\tt arXiv:1205.0176}}.

\bibitem{Duff:1990hn}
M.~J. Duff and J.~X. Lu, {\it {Duality Rotations in Membrane Theory}},  {\em
  Nucl.Phys.} {\bf B347} (1990) 394--419.

\bibitem{Duff:1989tf}
M.~J. Duff, {\it {Duality Rotations in String Theory}},  {\em Nucl.Phys.} {\bf
  B335} (1990) 610.

\bibitem{Tseytlin:1990nb}
A.~A. Tseytlin, {\it {Duality symmetric formulation of string world sheet
  dynamics}},  {\em Phys.Lett.} {\bf B242} (1990) 163--174.

\bibitem{Tseytlin:1990va}
A.~A. Tseytlin, {\it {Duality symmetric closed string theory and interacting
  chiral scalars}},  {\em Nucl.Phys.} {\bf B350} (1991) 395--440.

\bibitem{Wigner:1939}
E.~Wigner, {\it {On Unitary Representations of the Inhomogeneous Lorentz
  Group}},  {\em Annals Maths.} {\bf 40} (1939), no.~1 149--204.

\bibitem{Mackey:1951}
G.~W. Mackey, {\it {On Induced Representations of Groups}},  {\em Am.\ J.\
  Maths.} {\bf 73} (1951), no.~3 576--592.

\bibitem{Mackey:1952}
G.~W. Mackey, {\it {Induced Representations of Locally Compact Groups I}},
  {\em Annals Maths.} {\bf 55} (1952), no.~1 101--139.

\bibitem{Mackey:1955theory}
G.~W. Mackey, {\em {The Theory of Unitary Group Representations: Lecture Notes
  in Three Volumes}}.
\newblock Chicago Lectures in Mathematics. University of Chicago Press, 1955.

\bibitem{Mackey:2004mathematical}
G.~W. Mackey, {\em {The Mathematical Foundations of Quantum Mechanics: A
  Lecture-note Volume}}.
\newblock Mathematical physics monograph series. W.A. Benjamin, 1963.

\bibitem{Hull:1998vg}
C.~M. Hull, {\it {Timelike T duality, de Sitter space, large N gauge theories
  and topological field theory}},  {\em JHEP} {\bf 9807} (1998) 021,
  [\href{http://xxx.lanl.gov/abs/hep-th/9806146}{{\tt hep-th/9806146}}].

\bibitem{Hull:1998ym}
C.~M. Hull, {\it {Duality and the signature of space-time}},  {\em JHEP} {\bf
  9811} (1998) 017, [\href{http://xxx.lanl.gov/abs/hep-th/9807127}{{\tt
  hep-th/9807127}}].

\bibitem{Hull:1994ys}
C.~M. Hull and P.~K. Townsend, {\it {Unity of superstring dualities}},  {\em
  Nucl.\ Phys.} {\bf B 438} (1995) 109--137,
  [\href{http://xxx.lanl.gov/abs/hep-th/9410167}{{\tt hep-th/9410167}}].

\bibitem{Julia:1981sssg}
B.~{Julia}, {\it {Group Disintegrations}},  in {\em Superspace and
  Supergravity} (S.~W. {Hawking} and M.~{Rocek}, eds.), p.~331, 1981.

\bibitem{ThierryMieg:1981sssg}
J.~{Thierry-Mieg} and B.~{Morel}, {\it {Super Algebras in Exceptional
  Gravity}},  in {\em Superspace and Supergravity} (S.~W. {Hawking} and
  M.~{Rocek}, eds.), p.~351, 1981.

\bibitem{Isham:1971dv}
C.~J. Isham, A.~Salam, and J.~A. Strathdee, {\it {Nonlinear realizations of
  space-time symmetries. Scalar and tensor gravity}},  {\em Annals Phys.} {\bf
  62} (1971) 98--119.

\bibitem{Borisov:1974bn}
A.~B. Borisov and V.~I. Ogievetsky, {\it {Theory of Dynamical Affine and
  Conformal Symmetries as Gravity Theory}},  {\em Theor.\ Math.\ Phys.} {\bf
  21} (1975) 1179.

\bibitem{Andriot:2011uh}
D.~Andriot, M.~Larfors, D.~Lust, and P.~Patalong, {\it {A ten-dimensional
  action for non-geometric fluxes}},  {\em JHEP} {\bf 1109} (2011) 134,
  [\href{http://xxx.lanl.gov/abs/1106.4015}{{\tt arXiv:1106.4015}}].

\bibitem{Lunin:2005jy}
O.~Lunin and J.~M. Maldacena, {\it {Deforming field theories with U(1) x U(1)
  global symmetry and their gravity duals}},  {\em JHEP} {\bf 0505} (2005) 033,
  [\href{http://xxx.lanl.gov/abs/hep-th/0502086}{{\tt hep-th/0502086}}].

\bibitem{CatalOzer:2005mr}
A.~Catal-Ozer, {\it {Lunin-Maldacena deformations with three parameters}},
  {\em JHEP} {\bf 0602} (2006) 026,
  [\href{http://xxx.lanl.gov/abs/hep-th/0512290}{{\tt hep-th/0512290}}].

\bibitem{Frolov:2005dj}
S.~Frolov, {\it {Lax pair for strings in Lunin-Maldacena background}},  {\em
  JHEP} {\bf 0505} (2005) 069,
  [\href{http://xxx.lanl.gov/abs/hep-th/0503201}{{\tt hep-th/0503201}}].

\bibitem{Sen:1995cf}
A.~Sen, {\it {T duality of p-branes}},  {\em Mod.Phys.Lett.} {\bf A11} (1996)
  827--834, [\href{http://xxx.lanl.gov/abs/hep-th/9512203}{{\tt
  hep-th/9512203}}].

\bibitem{West:2010ev}
P.~C. West, {\it {$E_{11}$, generalised space-time and IIA string theory}},
  {\em Phys.Lett.} {\bf B696} (2011) 403--409,
  [\href{http://xxx.lanl.gov/abs/1009.2624}{{\tt arXiv:1009.2624}}].

\bibitem{Thompson:2011uw}
D.~C. Thompson, {\it {Duality Invariance: From M-theory to Double Field
  Theory}},  {\em JHEP} {\bf 1108} (2011) 125,
  [\href{http://xxx.lanl.gov/abs/1106.4036}{{\tt arXiv:1106.4036}}].

\bibitem{Witten:1995ex}
E.~Witten, {\it {String theory dynamics in various dimensions}},  {\em Nucl.\
  Phys.} {\bf B 443} (1995) 85--126,
  [\href{http://xxx.lanl.gov/abs/hep-th/9503124}{{\tt hep-th/9503124}}].

\bibitem{Schwarz:1995jq}
J.~H. Schwarz, {\it {The power of M theory}},  {\em Phys.\ Lett.} {\bf B 367}
  (1996) 97--103, [\href{http://xxx.lanl.gov/abs/hep-th/9510086}{{\tt
  hep-th/9510086}}].

\bibitem{Schwarz:1995dk}
J.~H. Schwarz, {\it {An SL(2,Z) multiplet of type IIB superstrings}},  {\em
  Phys.\ Lett.} {\bf B 360} (1995) 13--18,
  [\href{http://xxx.lanl.gov/abs/hep-th/9508143}{{\tt hep-th/9508143}}].

\bibitem{Schwarz:1996bh}
J.~H. Schwarz, {\it {Lectures on superstring and M theory dualities: Given at
  ICTP Spring School and at TASI Summer School}},  {\em Nucl.\ Phys.\ Proc.\
  Suppl.} {\bf 55B} (1997) 1--32,
  [\href{http://xxx.lanl.gov/abs/hep-th/9607201}{{\tt hep-th/9607201}}].

\bibitem{Vafa:1997pm}
C.~Vafa, {\it {Lectures on strings and dualities}},
  \href{http://xxx.lanl.gov/abs/hep-th/9702201}{{\tt hep-th/9702201}}.

\bibitem{Gutperle:2002ai}
M.~Gutperle and A.~Strominger, {\it {Space - like branes}},  {\em JHEP} {\bf
  0204} (2002) 018, [\href{http://xxx.lanl.gov/abs/hep-th/0202210}{{\tt
  hep-th/0202210}}].

\bibitem{Kruczenski:2002ap}
M.~Kruczenski, R.~C. Myers, and A.~W. Peet, {\it {Supergravity S-branes}},
  {\em JHEP} {\bf 0205} (2002) 039,
  [\href{http://xxx.lanl.gov/abs/hep-th/0204144}{{\tt hep-th/0204144}}].

\bibitem{Deger:2011nb}
N.~S. Deger and A.~Kaya, {\it {Deformations of Cosmological Solutions of D=11
  Supergravity}},  {\em Phys.Rev.} {\bf D84} (2011) 046005,
  [\href{http://xxx.lanl.gov/abs/1104.4019}{{\tt arXiv:1104.4019}}].

\bibitem{Gueven:1992hh}
R.~Gueven, {\it {Black p-brane solutions of D = 11 supergravity theory}},  {\em
  Phys.\ Lett.} {\bf B 276} (1992) 49--55.

\bibitem{Page:1984qv}
D.~N. Page, {\it {Classical stability of round and squashed seven spheres in
  eleven-dimensional supergravity}},  {\em Phys.\ Rev.} {\bf D 28} (1983) 2976.

\bibitem{Harrison:1968}
B.~K. Harrison, {\it {New Solutions of the Einstein-Maxwell Equations from
  Old}},  {\em J.\ Math.\ Phys.} {\bf 9} (1968) 1744--1752.

\bibitem{Duff:1990xz}
M.~J. Duff and K.~S. Stelle, {\it {Multimembrane solutions of D = 11
  supergravity}},  {\em Phys.\ Lett.} {\bf B 253} (1991) 113--118.

\bibitem{Hull:1998br}
C.~M. Hull and B.~Julia, {\it {Duality and moduli spaces for timelike
  reductions}},  {\em Nucl.\ Phys.} {\bf B 534} (1998) 250--260,
  [\href{http://xxx.lanl.gov/abs/hep-th/9803239}{{\tt hep-th/9803239}}].

\end{thebibliography}\endgroup

\end{document}